\documentclass[preprint,prx,amsmath,amssymb,longbibliography,superscriptaddress]{revtex4-1}

\pdfoutput=1

\usepackage{gensymb}
\usepackage{graphicx}
\usepackage{amsmath,latexsym,tabularx}
\usepackage{xcolor}
\usepackage{multirow}
\usepackage[
  colorlinks=true,
  urlcolor=blue,
  linkcolor=blue,
  citecolor=blue
]{hyperref}
\usepackage{multirow,booktabs}
\usepackage{float}

\newcommand{\affilLL}[0]{Lincoln Laboratory, Massachusetts Institute of Technology, Lexington, Massachusetts 02421, USA}
\newcommand{\affilMIT}{Massachusetts Institute of Technology, Cambridge, Massachusetts 02139, USA}

\newcommand{\specificthanks}[1]{\@fnsymbol{#1}}

\begin{document}

\title{Cooling of an Integrated Brillouin Laser below the Thermal Limit}

\author{William Loh}
%\thanks{Thanks.}
\affiliation{\affilLL}

\author{Dave Kharas}
\affiliation{\affilLL}

\author{Ryan Maxson}
 Correspondence to: William.Loh@ll.mit.edu
\affiliation{\affilLL}

\author{Gavin N. West}
\affiliation{\affilMIT}

\author{Alexander Medeiros}
\affiliation{\affilLL}

\author{Danielle Braje}
\affiliation{\affilLL}

\author{Paul W. Juodawlkis}
\affiliation{\affilLL}

\author{Robert McConnell}
%\email[]{robert.mcconnell@ll.mit.edu}
\affiliation{\affilLL}

\begin{abstract}

Photonically integrated resonators are promising as a platform for enabling ultranarrow linewidth lasers in a compact form factor. Owing to their small size, these integrated resonators suffer from thermal noise that limits the frequency stability of the optical mode to $\thicksim$100 kHz. Here, we demonstrate an integrated stimulated Brillouin scattering (SBS) laser based on a large mode-volume annulus resonator that realizes an ultranarrow thermal-noise-limited linewidth of 270 Hz. In practice, yet narrower linewidths are required before integrated lasers can be truly useful for applications such as optical atomic clocks, quantum computing, gravitational wave detection, and precision spectroscopy. To this end, we employ a thermorefractive noise suppression technique utilizing an auxiliary laser to reduce our SBS laser linewidth to 70 Hz. This demonstration showcases the possibility of stabilizing the thermal motion of even the narrowest linewidth chip lasers to below 100 Hz, thereby opening the door to making integrated microresonators practical for the most demanding future scientific endeavors.

\end{abstract}

\maketitle

%%%%%%%%%%%%%%%%%%%%%%%%%%%%%%%%%%%%%%%%%%%%%%%%%%%%%%%%%%%%%%%%

%\section{Introduction}

The ability to reach Hertz-class laser linewidths on chip is one of the key advances necessary to enabling portability in a number of scientific apparatuses that today only reside in laboratory settings \cite{Hinkley2013, Bloom2014, Godun2014, Huntemann2016, Koller2017, Brewer2019, Cirac1995, Abramovici1992, Abbott2016, Rafac2000}. Previously chip-based lasers have showcased the potential for fundamental linewidths of a few Hertz or below, derived from the lasers' measured white frequency noise at high offset frequencies \cite{Lee2012, Santis2014, Morton2018, Gundavarapu2019, Jin2021}. However, this performance does not typically translate to lower offset frequencies, which results in degradation of the lasers' integrated linewidths to values of 10 kHz or broader. Despite the substantial advantages of noise filtering offered by the Brillouin gain medium \cite{Debut2000}, prior standalone chip-based SBS lasers have also only reached integrated linewidths of 2-3 kHz. These SBS lasers have, however, been locked to secondary external-cavity microrod resonators \cite{Loh2015} and integrated microcavities \cite{Liu2021} to attain impressive linewidths of 95 Hz and 330 Hz, respectively, albeit at the cost of optical power, additional complexity, and twice the number of components needed to accomplish Pound-Drever-Hall (PDH) stabilization \cite{Drever1983}.  A singular resonator solution would pave the way for making integrated resonators practical to use over bulk or fiber-optic cavities, especially when size, weight, and power are of importance.

In regards to generating ultranarrow linewidth lasers, thermorefractive noise \cite{Gorodetsky2004, Notcutt2006, Matsko2007, Webster2008, Sun2017, Lim2019, Huang2019, Panuski2020, Jin2021} is especially pernicious as it causes frequency motion of the resonances in an optical cavity. This motion directly transfers to any laser created from the resonator and sets a bound on the minimum linewidth achievable. The variance of thermorefractive temperature fluctuations $<\delta T^2>$ is governed by \cite{Gorodetsky2004, Huang2019} \begin{equation} \langle\delta T^2\rangle = \frac{k_BT^2}{\rho CV} \end{equation}
where $k_B$ is Boltzmann's constant, $T$ is the temperature, $\rho$ is the density of the medium, $C$ is the specific heat, and $V$ is the volume. Unfortunately, the choice of material system used to realize ultrahigh quality factor (Q) resonators, while necessary to efficiently excite the SBS nonlinearity, also restricts many of the parameters available for reducing the magnitude of thermal fluctuations. The main accessible parameter is the resonator volume, which is physically limited to the size of the semiconductor chip. Recently, 1.41-m long spiral resonators have been demonstrated to suppress thermorefractive noise fluctuations and to dramatically quench laser noise at low offset frequencies \cite{Li2021}. Here, we explore the use of a large mode volume 20.5-mm diameter annulus resonator which exhibits low intrinsic temperature fluctuations. We utilize this resonator in the demonstration of an integrated SBS laser \cite{Grudinin2009, Lee2012, Kabakova2013, Loh2015, Otterstrom2018, Gundavarapu2019, Geng2006, Shee2011, Tow2012, Debut2000, Loh2019, Loh2020} that exhibits an ultralow thermorefractive-noise-limited linewidth of 270 Hz on chip.

%\section{Figure 1 Discussion}

\begin{figure}[t b !]
\includegraphics[width = 0.9 \columnwidth]{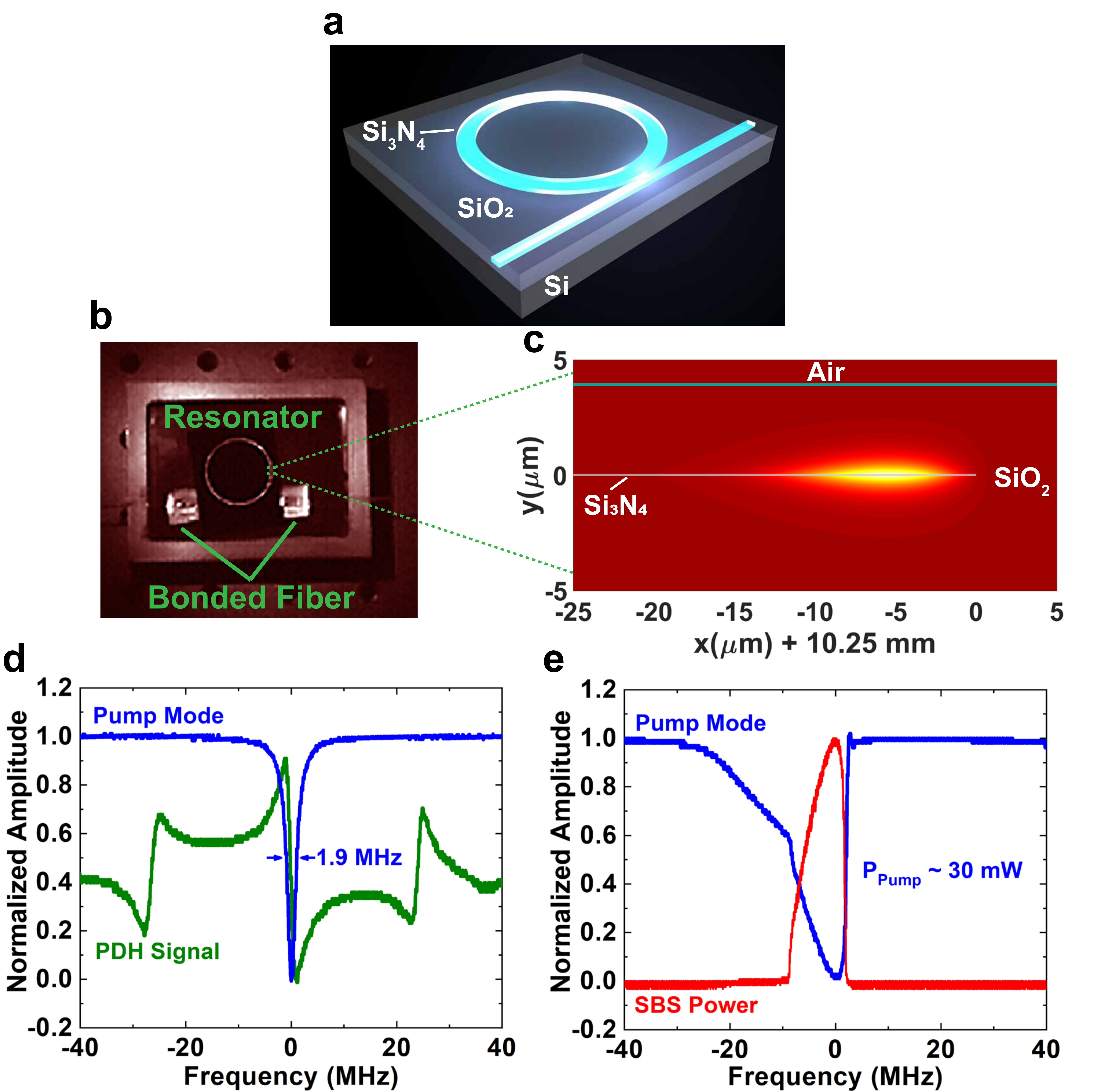}
\caption{
    \textbf{SBS Annulus Resonator and Mode Profile.}
    \textbf{a}, Illustration of the SBS resonator consisting of a thin layer of Si$_3$N$_4$  buried in SiO$_2$. 
    \textbf{b}, Infrared photograph of the SBS resonator bonded to two fiber couplers. The integrated chip is packaged in a 3 $\times$ 2.25 $\times$ 1 inch  copper enclosure.
    \textbf{c}, Simulated mode profile of the SBS annulus resonator. Only the peripheral of the 100 $\mu$m Si$_3$N$_4$ width where the mode is confined is shown. 
    \textbf{d}, Measured lineshape from a pump laser scan over a resonator mode performed at low optical powers. The mode is critically coupled and exhibits a loaded linewidth of 1.9 MHz. The PDH scan with 25 MHz sideband spacing is also superimposed.
    \textbf{e}, Scan of the pump laser over the resonator mode at high optical powers. The backreflected SBS power is also plotted. The SBS power builds up to a maximum at resonance.
}
\label{fig:fig1}
\end{figure}

Figure 1a depicts a schematic of our SBS laser operating at 1348 nm wavelength, which comprises a 40-nm thick Si$_3$N$_4$ annulus resonator encapsulated by SiO$_2$. The choice of operation at 1348 nm was motivated by ease of accessing the $^{88}$Sr$^+$ ion at 674 nm via frequency doubling for the application of an optical atomic clock. The configuration of our resonator forms a diffuse optical mode \cite{Bauters2011} that is primarily overlapped with the low-optical loss oxide. We use a ring waveguide width of 100 um, forming an annulus resonator. This distances the inner Si$_3$N$_4$ sidewall from the optical mode, thereby reducing scattering loss, while also constraining the significant stress that would otherwise be induced by a 20.5-mm layer of Si$_3$N$_4$. An infrared photograph of the SBS resonator is shown in Fig. 1b with the resonator illuminated and bonded to input and output coupling fibers. The annulus itself appears significantly brighter than the bus waveguide due to the power enhancement exhibited by the ultralow-loss resonator. The SBS resonator is packaged in a copper enclosure that serves to isolate the chip from external vibrations and temperature fluctuations. The simulated optical mode profile (Fig. 1c) is asymmetric with 99$\%$ of the optical energy contained within a width of 19.1 $\mu$m and a height of 4.6 $\mu$m.

Figure 1d shows a scan of the pump laser over the cavity resonance at low optical powers where thermal effects that broaden the resonance are insignificant. The resonance exhibits a linewidth of 1.9 MHz, which for a 1348 nm optical carrier corresponds to an intrinsic Q of 230 million and an intracavity loss of 0.1 dB/m. The PDH error signal used for locking the pump to resonance is also superimposed onto the resonance lineshape. At the high pump powers necessary for the generation of SBS (Fig. 1e), the resonance lineshape exhibits significant thermal asymmetry. The backwards propagating SBS power is superimposed on the resonance scan which shows the build up of SBS power as the pump power is coupled into the cavity.

%\section{Figure 2 Discussion}

\begin{figure}[t b !]
\includegraphics[width = 0.8 \columnwidth]{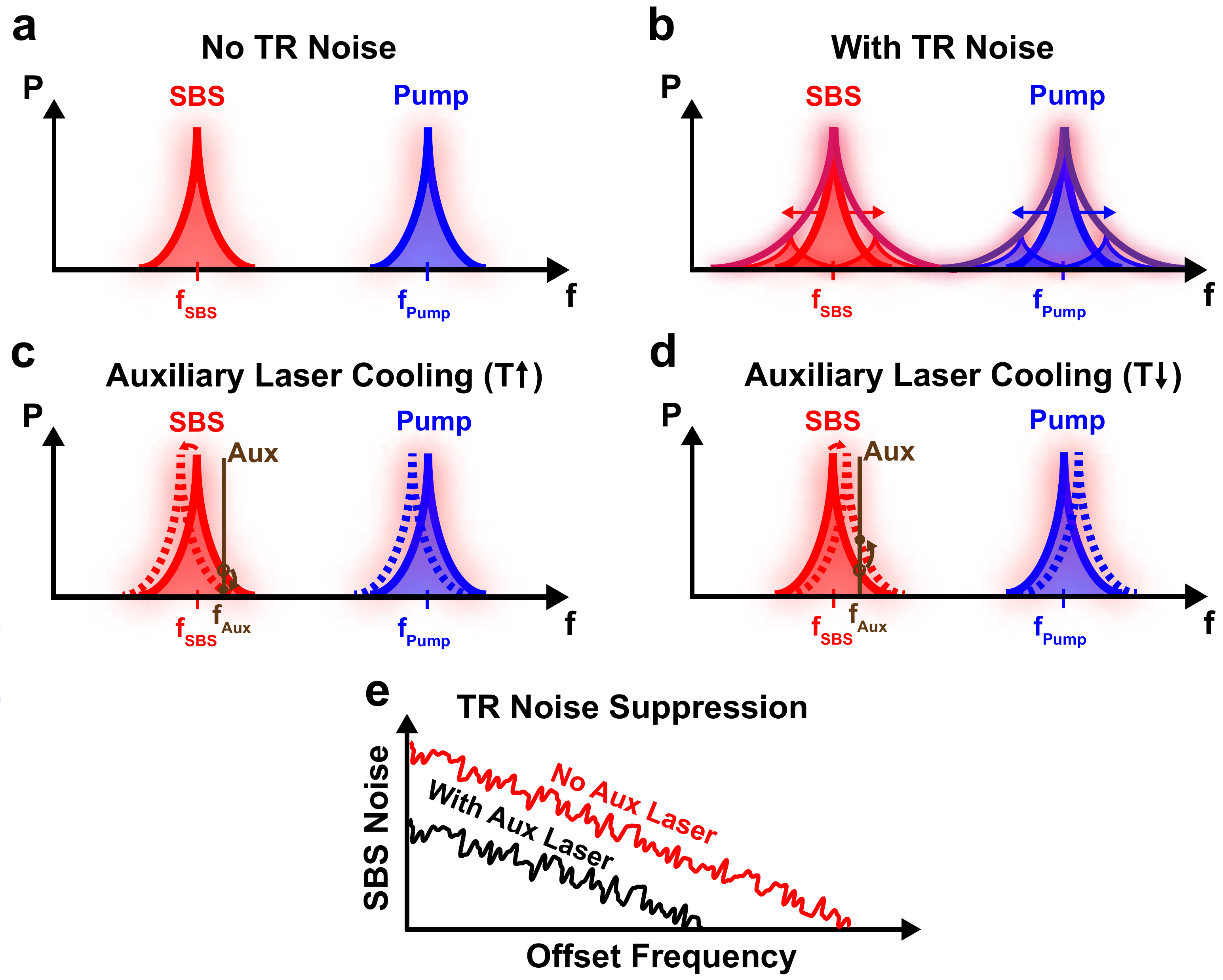}
\caption{
    \textbf{Concept of Auxiliary laser thermorefractive noise stabilization.}
    \textbf{a}, Illustration of two resonator modes (SBS and Pump) used for operating an SBS laser. Without the presence of thermorefractive (TR) noise, the two modes each exhibit a linewidth determined by the resonator optical loss.
    \textbf{b}, Diagram of the two modes with thermorefractive noise present. The modes shift in frequency due to random temperature fluctuations, which broadens the effective mode linewidth
    \textbf{c}, Illustration of the use of an auxiliary (Aux) laser to stabilize the thermorefractive noise. The auxiliary laser is placed on the side of the SBS resonance to counteract the thermal motion of the resonator mode. The resonator modes decreases in frequency when the temperature increases, while the independent auxiliary laser remains fixed in place.
    \textbf{d}, Case for a decrease in temperature causing the resonator modes to increase in frequency. The position of the mode changes relative to that of the auxiliary laser and stabilizes the cavity temperature.
    \textbf{e}, Representation of the frequency noise of the stabilized SBS laser. The thermorefractive noise imprinted on the SBS laser is suppressed by the auxiliary laser.
}
\label{fig:fig2}
\end{figure}

As our current resonator diameter of 20.5 mm spans nearly the entire fabrication reticle, we look for additional means of narrowing its linewidth beyond just increasing mode volume. Recently, it was shown that the thermal noise of a microresonator frequency comb could be stabilized through the use of an independent auxiliary laser coupled into the cavity. The coupled power of the auxiliary laser induces a counteracting thermal force that reduced the linewidth of the carrier-envelope offset beat from 2.2 MHz to 280 kHz \cite{Sun2017, Drake2020}. We demonstrate an analogous method here for our case of a Hertz-class linewidth SBS laser, utilizing an auxiliary laser to dampen the thermal motion of the optical mode. Figure 2 illustrates this concept in greater detail. The presence of thermorefractive noise causes the two modes (pump and SBS, Fig. 2a) used in operating our SBS laser to fluctuate in frequency and to broaden (Fig. 2b). By placing an auxiliary laser slightly detuned from a resonator mode, any temperature-induced resonance shift results in a change in the laser power coupled into and circulating within the resonator (Figs. 2c and 2d). Through absorptive self-heating, this change in circulating power generates a second shift opposite to the initial temperature fluctuation, thus damping the overall resonance motion and effectively \textquotedblleft cooling" thermal fluctuations. When the two resonances are then used for SBS generation, the thermorefractive noise suppression transfers over to the resulting SBS signal (Fig. 2e). Although we have referred to this technique as cooling, we note that the resonator itself does not physically reduce in temperature. Rather, the resonator takes on a linewidth with reduced thermal broadening that is characteristic of a lower operating temperature.

%\section{Figure 3 Discussion}

\begin{figure}[t b !]
\includegraphics[width = 1.0 \columnwidth]{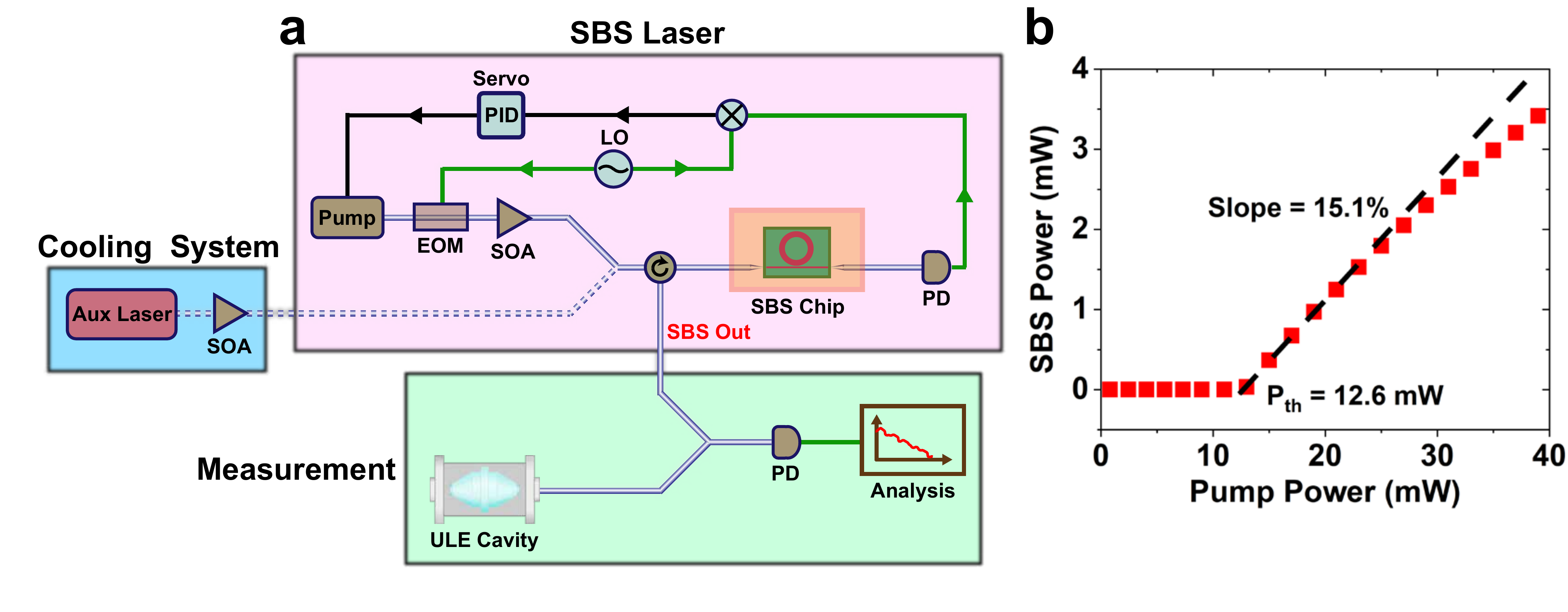}
\caption{
    \textbf{Chip SBS Laser System and Operation.}
    \textbf{a}, Diagram of the SBS system comprising a pump laser that excites a packaged SBS resonator. An electro-optic modulator (EOM), semiconductor optical amplifier (SOA), photodiode (PD), local oscillator (LO) and proportional-integral-derivative (PID) controller all enable the pump to be PDH locked to the cavity resonance. An auxiliary (Aux) laser is used to stabilize the thermorefractive noise of the SBS cavity. An ultralow expansion (ULE) cavity serves as a reference for the analysis of the SBS laser noise.
    \textbf{b}, SBS power out for an applied pump power. The powers are reported at the SBS chip interface. The threshold power measured is 12.6 mW, and the slope efficiency is 15.1 $\%$.
}
\label{fig:fig3}
\end{figure}

Our SBS system (Fig. 3a) comprises three separate elements: the SBS laser itself, the measurement system, and the auxiliary laser system used for cooling the resonator's thermorefractive noise. Our basic SBS laser consists of a pump laser that is amplified and sent into a packaged integrated ultrahigh-Q resonator to excite the SBS nonlinearity. The pump is phase modulated and photodetected post-resonator to PDH lock the pump light to the cavity resonance. The ultranarrow linewidth SBS light is generated in the counterpropagating direction and is coupled out of a circulator. The integrated SBS laser produces 1.1 mW of output power for 20 mW of pump input (Fig. 3b), which yields a slope efficiency of 15.1 $\%$ and a threshold pump power (P$_{th}$) of 12.6 mW. These measurements include the losses incurred by both the pump and SBS lasers upon coupling into and out of the SBS chip. By measuring the total loss of the SBS chip and assuming the waveguide losses to be negligible, we determine the input chip coupling loss to be 2.1 dB. Accounting for this coupling loss, the slope efficiency of the SBS laser with ideal coupling is 40 $\%$. The pump threshold is 7.8 mW in this ideal case, which would make the chip performance competitive with that of fiber SBS lasers \cite{Loh2019}.

In the auxiliary laser system, a secondary input port allows for an auxiliary laser to be coupled into the SBS resonator in order to stabilize the resonator's thermal fluctuations. It is imperative that this auxiliary laser is exceptionally stable to prevent the unintentional transfer of the auxiliary laser noise into the cavity. This noise transfer occurs via the inverse process of the resonator cooling where the frequency motion of the auxiliary laser now causes motion of the cavity resonance via an induced temperature change. For our experiments here, we use an ultranarrow 20-Hz linewidth fiber SBS laser as the auxiliary cooling laser \cite{Loh2019, Loh2020}.

Finally, in our measurement system, the SBS output is combined with light derived from a ULE reference cavity, and their heterodyne beat is analyzed to characterize the performance of the integrated SBS laser. Through these measurements, we extract estimates for the linewidth of the SBS laser with and without the cooling laser applied. From the frequency noise of the SBS laser (Fig. 4a), we find that the noise spectral density reaches 1400 Hz$^2$/Hz at 10 Hz offset frequency and decreases to $<$ 1.4 Hz$^2$/Hz beyond 100 kHz offset. By integrating the spectral density of noise \cite{Hjelme1991}, we determine the integrated linewidth of the SBS laser to be 210 Hz. After locking to the cavity resonance, the pump laser also exhibits a comparable level of frequency noise performance for offset frequencies below 1 kHz. This frequency noise level is set by thermorefractive noise, as the motion of the cavity resonance transfers directly to both the cavity-locked pump and the output SBS signals. At higher frequencies where the servo bandwidth cannot respond, the pump noise rises above the noise of the SBS laser.

Upon application of the auxiliary laser (Fig. 4b), the frequency noise of the SBS laser significantly decreases and reaches 90 Hz$^2$/Hz at 10 Hz offset. This decrease in noise corresponds to cooling of the SBS laser to an effective temperature of 73 K, reaching an integrated linewidth of 60 Hz. The stabilized pump laser, on the other hand, only decreases slightly in noise after the auxiliary laser is applied. The pump instead becomes limited by the noise floor of the PDH lock, which is only slightly below the level of thermorefractive noise. The SBS suppression of the pump laser noise becomes evident upon comparison of the pump and SBS frequency noise after cooling. The observed suppression is 10$\times$, which is substantially lower than the expected suppression factor of $>$1000$\times$. Comparing the cooled SBS laser to the auxiliary laser's frequency noise, we observe that the correction limit is only reached at low offset frequencies. For offset frequencies above 10 Hz, the cooled SBS laser noise gradually rises above the correction limit, which we attribute to be due to a rolloff in the thermal response \cite{Sun2017}.

The lineshape of the SBS laser (Fig. 4c) provides additional confirmation of the benefits of auxiliary laser cooling. The measured 3-dB full width half maximum (FWHM) linewidth of the SBS laser initially starts at a measured value of 270 Hz. Upon application of the auxiliary laser, the SBS laser reaches a linewidth of 70 Hz. These values correspond well to the extracted values of 210 Hz and 60 Hz found by integrating the laser frequency noise \cite{Hjelme1991}.

%\section{Figure 4 Discussion}

\begin{figure}[t b !]
\includegraphics[width = 0.61 \columnwidth]{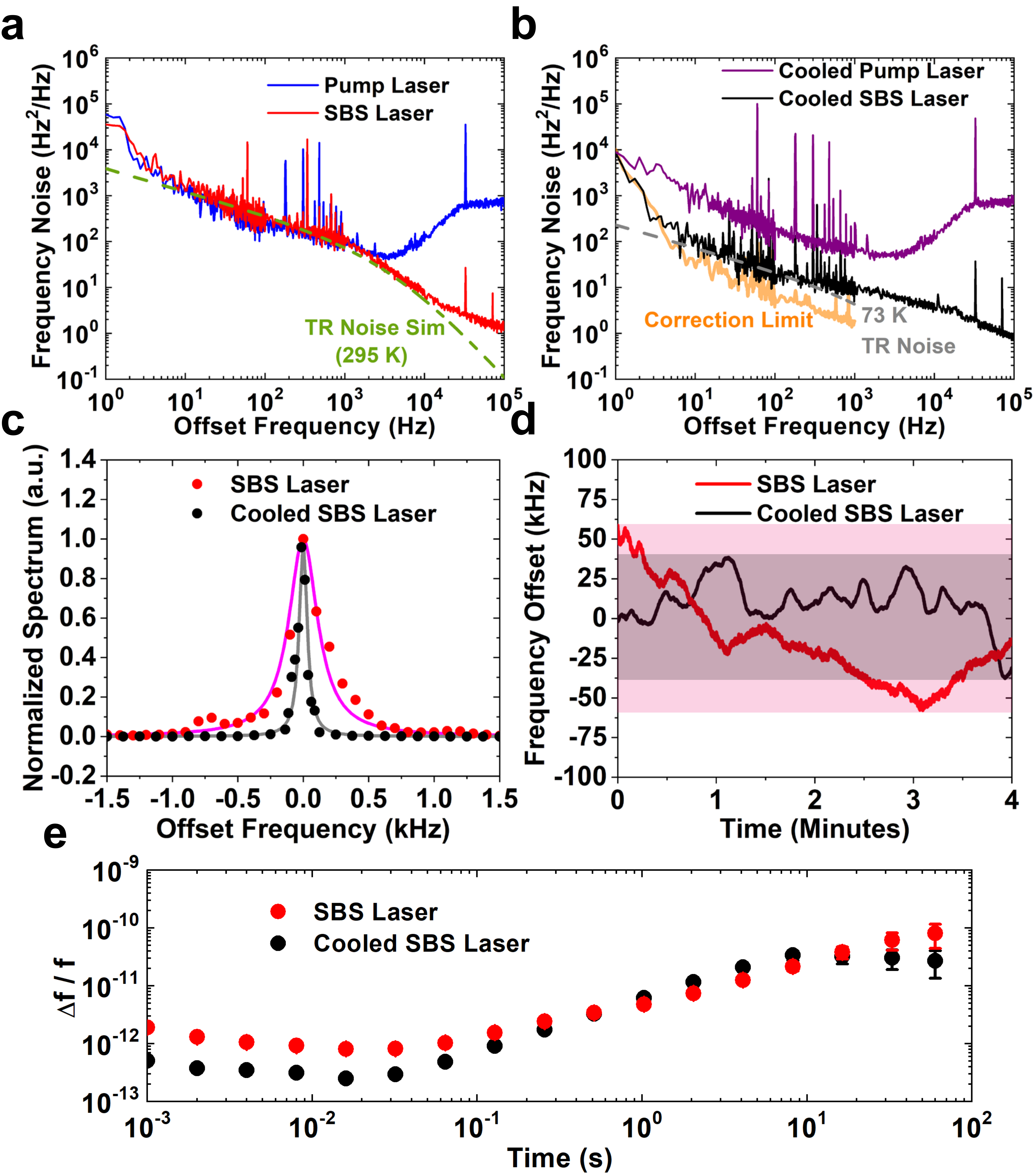}
\caption{
    \textbf{Measurements of the SBS laser.}
    \textbf{a}, Frequency noise of the SBS laser (red) and of the stabilized pump laser (blue). At lower offset frequencies, both lasers reach the 295 K thermorefractive (TR) noise limit (dashed green).
    \textbf{b}, Measured frequency noise when the auxiliary laser is turned on. Compared to the SBS laser from (a), the cooled SBS laser (black) exhibits significantly lower noise and a reduced effective temperature from 295 K to 73 K (dashed gray). The stabilized pump laser (purple) experiences only a slight reduction in noise. The correction limit determined by the noise floor of the auxiliary laser (orange) is also shown for frequencies up to the resonator's thermal response.
    \textbf{c}, Lineshape of the SBS laser before and after auxiliary laser cooling. The resolution bandwidth used is 18.75 Hz, and the sweep time is 80 ms. The SBS laser linewidth is reduced by a factor of 4$\times$ after cooling.
    \textbf{d}, Time series traces of the SBS laser with (black) and without (red) the cooling laser present. The cooled SBS laser exhibits significantly lower short-term frequency fluctuations. However, its drift over the 4 minute time span is only slightly reduced from that of the standard SBS laser.
    \textbf{e}, Fractional frequency fluctuations of the SBS laser with (black) and without (red) auxiliary laser cooling.
}
\label{fig:fig4}
\end{figure}

It is interesting to examine the effects of the cooling laser on the SBS laser's long-term noise. From the measured time series (Fig. 4d), the cooling clearly reduces the fast fluctuations of the SBS laser frequency. However, over the time scale of minutes, the effect of the cooling only has a mild effect on reducing long-term drift. At these long time scales, the cooled SBS laser follows the drift of the auxiliary laser (see Fig. 4b), which is only slightly lower than that of the integrated SBS laser as a consequence of the high degree to which environmental noise is mitigated in both laser systems. The fractional frequency noise of the SBS laser (Fig. 4e) exhibits a minimum noise of 8$\times$10$^{-13}$ near 15 ms averaging time when no auxiliary laser is applied. However, with the auxiliary laser cooling the cavity, the SBS laser noise reduces to 2.5$\times$10$^{-13}$ near 15 ms. At the carrier frequency of 222.5 THz, these values correspond to frequency fluctuations of 180 Hz and 56 Hz, respectively for the standard and cooled cases, and also agree with the measured linewidths extracted from Figs. 4b and c. Over long time scales, the fractional frequency noise of the two lasers begin to converge in value owing to the similar level of laser drift.

%\section{Conclusion}

Recently, fiber-based SBS lasers have demonstrated exceptionally low noise in a compact form factor, now reaching the levels needed to operate advanced scientific instrumentation such as optical atomic clocks \cite{Loh2020}. A truly portable system, however, will benefit from full integration of all of its components, including the SBS resonator and its surrounding laser infrastructure. Integrated lasers have achieved Hertz-class fundamental linewidths, but have seen this performance rapidly degrade at longer time scales where most real-world applications operate. Significant progress has been made recently through the use of chip SBS lasers and stabilizing these lasers to external microrod \cite{Loh2015} and microresonator \cite{Liu2021} cavities, but at a substantial cost in system complexity. Our SBS laser shown here uses a single integrated annulus resonator to generate an ultranarrow 270-Hz linewidth light source. We further advance the understanding of thermorefractive noise in microcavities by applying a stabilization procedure to cool the thermal motion of our SBS resonator. This technique reduces the SBS laser linewidth to 70 Hz but requires the use of an additional stable auxiliary laser separate to the main SBS laser system. Future improvements may uncover paths to self-cooling using the SBS output itself without necessitating the use of an auxiliary laser source. This would be possible, for example, by using the SBS output to probe a second resonance that has a different effective thermo-optic coefficient. Frequency doubling of the SBS output, which is already required for addressing relevant atomic transitions, would be one way to accomplish this as the resonance behavior is expected to be substantially different at the harmonic frequency.

\clearpage

\section{Methods}

\subsection{SBS Laser Fabrication}

The SBS resonators were fabricated at MIT Lincoln Laboratory's Microelectronics Laboratory (ML), a 200-mm CMOS capable fabrication facility. Processing started with 200-mm silicon substrates prepared with a 15.0-$\mu$m thick layer of thermal oxide. A short 20 sec chemo-mechanical polish step (CMP) was performed on the oxide surface, enough to remove nominally $\sim$50 nm of oxide. This planarization step provides an oxide surface with nominally a 0.2--0.4 nm RMS roughness. After CMP, the wafers were cleaned and loaded for low pressure chemical vapor deposition (LPCVD) of a 40-nm thick stoichiometric silicon nitride film. Following deposition, the nitride film is annealed at 1100 $\degree$C in nitrogen for 3 hours. The SBS resonators are patterned in the nitride film using an ASML PAS 5500 193-nm stepper and etched using a LAM TCP 9400 etch system with a SF$_6$ based etch chemistry. After etching, the wafers are stripped of resist, cleaned, and overclad with a 4-$\mu$m thick LPCVD-deposited, TEOS (Tetraethyl orthosilicate) based silicon oxide film. To minimize the risk of cracking and to improve oxide optical quality, the TEOS oxide is deposited in 4 passes of 1 $\mu$m thickness each, with a 1100 $\degree$C 3 hour N$_2$ anneal after each pass.

Once oxide overcladding is completed, the dicing street region of the chip is etched to create an optically smooth facet to facilitate low loss coupling to the input nitride waveguide with an optical fiber block. The optical facet is created by patterning a street trench feature into a 16-$\mu$m thick resist, and then etching through the $\sim$19-$\mu$m thick oxide layer down to the silicon surface. With the resist in place, we utilize DRIE LAM Alliance system to deep etch an additional $\sim$500-$\mu$m deep silicon facet trench in the vicinity of the waveguide input, thus creating a smooth surface for subsequent fiber block attach. This trench does not penetrate the full 725 $\mu$m thickness of the wafer. At this point the wafer fabrication is complete and the wafers are coated with resist and brought to a packaging lab for dicing. After initial dicing, chips that are selected for fiber-block attach are re-mounted on tape, and then diced again on the back side to remove the remaining 225 $\mu$m-thick region around the input facet that was not etched earlier during the facet trench.

\subsection{SBS Resonator Packaging}

The SBS resonator is packaged by mounting the chip on a copper plate and affixing this plate within a copper enclosure. Single channel fiber arrays are aligned to the input and output waveguides and glued in place at the coupling interface. The alignment is performed by maximizing the power coupled through the device while maintaining the correct polarization orientation. An additional encapsulant glue is then flowed over the fiber to further protect against potential future shifts in coupling. The fibers exit the copper enclosure through two miniature holes and serve as the only interface to the resonator from the outside. In order to limit the SBS drift at long time scales, the copper enclosure is temperature controlled to within +/- 0.1 mK via a thermistor.

\subsection{SBS Laser Measurement and Thermorefractive Noise Suppression}

We suppress the thermorefractive noise by injecting a 20 Hz linewidth fiber SBS laser into the SBS chip resonator. The auxiliary laser is sent in the forward direction along with the pump light so that the extracted SBS light in the reverse-propagating direction could be isolated with a circulator. The auxiliary laser is tuned into the cavity resonance and reaches a thermally stable point along the resonance slope. The power sent into the SBS resonator was varied by a factor of two in the range of 5 mW, however, we did not find the thermorefractive noise suppression to be particularly sensitive to auxiliary laser power. This is in contrast to the results reported in Ref. \cite{Drake2020} where the thermorefractive noise suppression increased with increasing auxiliary laser power. We attribute these differences to our cavity Q being a factor of $100\times$ larger, which results in a $10,000\times$ increase in the resonance slope. This increase in slope results in a $\sim10,000\times$ stronger thermal response of the cavity resonance to optical power and thus likely resulted in a saturation of the effects of optical power.

The SBS laser is measured by combining its output with the output of a ULE reference cavity and heterodyning the result together on a photodiode. The reference cavity exhibits a fractional frequency noise of 3$\times$10$^{-15}$ at 1 second averaging time. Since the noise of the reference cavity is much lower than that of the chip SBS laser, the heterodyne microwave output will be a faithful representation of the SBS laser noise. We measure the frequency noise of the SBS laser by sending the microwave output through a frequency-to-voltage converter and processing the output on a low-frequency spectrum analyzer. For offset frequencies beyond 10 kHz, the frequency noise is measured via an unbalanced Mach-Zehnder delay line with 250 m delay \cite{Llopis2011}. The results are combined to produce a coherent frequency noise spectrum. The SBS laser's RF spectrum is measured by directly putting the heterodyne output on a microwave spectrum analyzer. The fractional frequency noise is measured by sending the microwave output to a frequency counter with 1 ms gate time.

\subsection{Thermorefractive Noise Modelling}

Our simulations for thermorefractive noise are performed by assuming that nearly the entirety of the optical mode resides in SiO$_2$ and therefore only knowledge of the thermo-optic parameters pertaining to SiO$_2$ is required. We further assume operation at room temperature, leaving the details of the optical mode to be the primary unknown parameters needed in our modelling \cite{Huang2019}. The required mode spatial halfwidths could in principle be determined through mode simulations of the annulus resonator structure. However,the extracted values yielded predictions of thermorefractive-limited frequency noise that were too low and also noise corner frequencies that did not agree with our measured results. By fitting the mode halfwidths, we found excellent agreement in both the absolute noise magnitude and also the corner frequency when the mode halfwidths were 70$ \%$ of their simulated value. We attribute these differences to our dilute SBS mode, which renders the exact mode shape and size to be highly dependent on small changes of refractive index. Our modelled results are plotted alongside the measurement in Figs. 4(a) and (b).

%%%%%%%%%%%%%%%%%%%%%%%%%%%%%%%%%%%%%%%%%%%%%%%%%%%%%%%%%%%%%%%%

\section{Data availability}

The data sets that support this study are available on reasonable request.

\section{Code availability}

The code used for analysis and simulations are available on reasonable request.

%%%%%%%%%%%%%%%%%%%%%%%%%%%%%%%%%%%%%%%%%%%%%%%%%%%%%%%%%%%%%%%%

\section{Acknowledgements}

We thank J. Chiaverini, S. Yegnanarayanan, A. Libson, and C. Sorace-Agaskar for helpful discussions.

DISTRIBUTION STATEMENT A. Approved for public release. Distribution is unlimited. This material is based upon work supported by the Under Secretary of Defense for Research and Engineering under Air Force Contract No. FA8702-15-D-0001. Any opinions, findings, conclusions or recommendations expressed in this material are those of the author(s) and do not necessarily reflect the views of the Under Secretary of Defense for Research and Engineering.

\section{Contributions}

W.L. and G.N.W. conceived and designed the experiments. D.K. performed the fabrication of the chips, and W.L. performed the testing. R.M., A.M., and W.L. performed the packaging and bonding of the SBS chips. All authors discussed the results and contributed to the manuscript.

\section{Competing interests}

The authors declare no competing financial interests.

%%%%%%%%%%%%%%%%%%%%%%%%%%%%%%%%%%%%%%%%%%%%%%%%%%%%%%%%%%%%%%%%

\bibliography{SBS_Chip}

\begin{thebibliography}{40}%
\makeatletter
\providecommand \@ifxundefined [1]{%
 \@ifx{#1\undefined}
}%
\providecommand \@ifnum [1]{%
 \ifnum #1\expandafter \@firstoftwo
 \else \expandafter \@secondoftwo
 \fi
}%
\providecommand \@ifx [1]{%
 \ifx #1\expandafter \@firstoftwo
 \else \expandafter \@secondoftwo
 \fi
}%
\providecommand \natexlab [1]{#1}%
\providecommand \enquote  [1]{``#1''}%
\providecommand \bibnamefont  [1]{#1}%
\providecommand \bibfnamefont [1]{#1}%
\providecommand \citenamefont [1]{#1}%
\providecommand \href@noop [0]{\@secondoftwo}%
\providecommand \href [0]{\begingroup \@sanitize@url \@href}%
\providecommand \@href[1]{\@@startlink{#1}\@@href}%
\providecommand \@@href[1]{\endgroup#1\@@endlink}%
\providecommand \@sanitize@url [0]{\catcode `\\12\catcode `\$12\catcode
  `\&12\catcode `\#12\catcode `\^12\catcode `\_12\catcode `\%12\relax}%
\providecommand \@@startlink[1]{}%
\providecommand \@@endlink[0]{}%
\providecommand \url  [0]{\begingroup\@sanitize@url \@url }%
\providecommand \@url [1]{\endgroup\@href {#1}{\urlprefix }}%
\providecommand \urlprefix  [0]{URL }%
\providecommand \Eprint [0]{\href }%
\providecommand \doibase [0]{http://dx.doi.org/}%
\providecommand \selectlanguage [0]{\@gobble}%
\providecommand \bibinfo  [0]{\@secondoftwo}%
\providecommand \bibfield  [0]{\@secondoftwo}%
\providecommand \translation [1]{[#1]}%
\providecommand \BibitemOpen [0]{}%
\providecommand \bibitemStop [0]{}%
\providecommand \bibitemNoStop [0]{.\EOS\space}%
\providecommand \EOS [0]{\spacefactor3000\relax}%
\providecommand \BibitemShut  [1]{\csname bibitem#1\endcsname}%
\let\auto@bib@innerbib\@empty
%</preamble>
\bibitem [{\citenamefont {Hinkley}\ \emph {et~al.}(2013)\citenamefont
  {Hinkley}, \citenamefont {Sherman}, \citenamefont {Philips}, \citenamefont
  {Schioppo}, \citenamefont {Lemke}, \citenamefont {Beloy}, \citenamefont
  {Pizzocaro}, \citenamefont {Oates},\ and\ \citenamefont
  {Ludlow}}]{Hinkley2013}%
  \BibitemOpen
  \bibfield  {author} {\bibinfo {author} {\bibfnamefont {N.}~\bibnamefont
  {Hinkley}}, \bibinfo {author} {\bibfnamefont {J.~A.}\ \bibnamefont
  {Sherman}}, \bibinfo {author} {\bibfnamefont {N.~B.}\ \bibnamefont
  {Philips}}, \bibinfo {author} {\bibfnamefont {M.}~\bibnamefont {Schioppo}},
  \bibinfo {author} {\bibfnamefont {N.~D.}\ \bibnamefont {Lemke}}, \bibinfo
  {author} {\bibfnamefont {K.}~\bibnamefont {Beloy}}, \bibinfo {author}
  {\bibfnamefont {M.}~\bibnamefont {Pizzocaro}}, \bibinfo {author}
  {\bibfnamefont {C.~W.}\ \bibnamefont {Oates}}, \ and\ \bibinfo {author}
  {\bibfnamefont {A.~D.}\ \bibnamefont {Ludlow}},\ }\bibfield  {title}
  {\enquote {\bibinfo {title} {An atomic clock with $10^{-18}$ instability},}\
  }\href {\doibase 10.1126/science.1240420} {\bibfield  {journal} {\bibinfo
  {journal} {Science}\ }\textbf {\bibinfo {volume} {341}},\ \bibinfo {pages}
  {1215--1218} (\bibinfo {year} {2013})}\BibitemShut {NoStop}%
\bibitem [{\citenamefont {Bloom}\ \emph {et~al.}(2014)\citenamefont {Bloom},
  \citenamefont {Nicholson}, \citenamefont {Williams}, \citenamefont
  {Campbell}, \citenamefont {Bishof}, \citenamefont {Zhang}, \citenamefont
  {Bromley},\ and\ \citenamefont {Ye}}]{Bloom2014}%
  \BibitemOpen
  \bibfield  {author} {\bibinfo {author} {\bibfnamefont {B.~J.}\ \bibnamefont
  {Bloom}}, \bibinfo {author} {\bibfnamefont {T.~L.}\ \bibnamefont
  {Nicholson}}, \bibinfo {author} {\bibfnamefont {J.~R.}\ \bibnamefont
  {Williams}}, \bibinfo {author} {\bibfnamefont {S.~L.}\ \bibnamefont
  {Campbell}}, \bibinfo {author} {\bibfnamefont {M.}~\bibnamefont {Bishof}},
  \bibinfo {author} {\bibfnamefont {X.}~\bibnamefont {Zhang}}, \bibinfo
  {author} {\bibfnamefont {S.~L.}\ \bibnamefont {Bromley}}, \ and\ \bibinfo
  {author} {\bibfnamefont {J.}~\bibnamefont {Ye}},\ }\bibfield  {title}
  {\enquote {\bibinfo {title} {An optical lattice clock with accuracy and
  stability at the $10^{-18}$ level},}\ }\href {\doibase 10.1038/nature12941}
  {\bibfield  {journal} {\bibinfo  {journal} {Nature}\ }\textbf {\bibinfo
  {volume} {506}},\ \bibinfo {pages} {71--75} (\bibinfo {year}
  {2014})}\BibitemShut {NoStop}%
\bibitem [{\citenamefont {Godun}\ \emph {et~al.}(2014)\citenamefont {Godun},
  \citenamefont {Nisbet-Jones}, \citenamefont {Jones}, \citenamefont {King},
  \citenamefont {Johnson}, \citenamefont {Margolis}, \citenamefont {Szymaniec},
  \citenamefont {Lea}, \citenamefont {Bongs},\ and\ \citenamefont
  {Gill}}]{Godun2014}%
  \BibitemOpen
  \bibfield  {author} {\bibinfo {author} {\bibfnamefont {R.~M.}\ \bibnamefont
  {Godun}}, \bibinfo {author} {\bibfnamefont {P.~B.~R.}\ \bibnamefont
  {Nisbet-Jones}}, \bibinfo {author} {\bibfnamefont {J.~M.}\ \bibnamefont
  {Jones}}, \bibinfo {author} {\bibfnamefont {S.~A.}\ \bibnamefont {King}},
  \bibinfo {author} {\bibfnamefont {L.~A.~M.}\ \bibnamefont {Johnson}},
  \bibinfo {author} {\bibfnamefont {H.~S.}\ \bibnamefont {Margolis}}, \bibinfo
  {author} {\bibfnamefont {K.}~\bibnamefont {Szymaniec}}, \bibinfo {author}
  {\bibfnamefont {S.~N.}\ \bibnamefont {Lea}}, \bibinfo {author} {\bibfnamefont
  {K.}~\bibnamefont {Bongs}}, \ and\ \bibinfo {author} {\bibfnamefont
  {P.}~\bibnamefont {Gill}},\ }\bibfield  {title} {\enquote {\bibinfo {title}
  {Frequency ratio of two optical clock transitions in $^{171}\mathrm{Yb}^{+}$
  and constraints on the time variation of fundamental constants},}\ }\href
  {\doibase 10.1103/PhysRevLett.113.210801} {\bibfield  {journal} {\bibinfo
  {journal} {Phys. Rev. Lett.}\ }\textbf {\bibinfo {volume} {113}},\ \bibinfo
  {pages} {210801} (\bibinfo {year} {2014})}\BibitemShut {NoStop}%
\bibitem [{\citenamefont {Huntemann}\ \emph {et~al.}(2016)\citenamefont
  {Huntemann}, \citenamefont {Sanner}, \citenamefont {Lipphardt}, \citenamefont
  {Tamm},\ and\ \citenamefont {Peik}}]{Huntemann2016}%
  \BibitemOpen
  \bibfield  {author} {\bibinfo {author} {\bibfnamefont {N.}~\bibnamefont
  {Huntemann}}, \bibinfo {author} {\bibfnamefont {C.}~\bibnamefont {Sanner}},
  \bibinfo {author} {\bibfnamefont {B.}~\bibnamefont {Lipphardt}}, \bibinfo
  {author} {\bibfnamefont {Chr.}\ \bibnamefont {Tamm}}, \ and\ \bibinfo
  {author} {\bibfnamefont {E.}~\bibnamefont {Peik}},\ }\bibfield  {title}
  {\enquote {\bibinfo {title} {Single-ion atomic clock with $3\times10^{-18}$
  systematic uncertainty},}\ }\href {\doibase 10.1103/PhysRevLett.116.063001}
  {\bibfield  {journal} {\bibinfo  {journal} {Phys. Rev. Lett.}\ }\textbf
  {\bibinfo {volume} {116}},\ \bibinfo {pages} {063001} (\bibinfo {year}
  {2016})}\BibitemShut {NoStop}%
\bibitem [{\citenamefont {Koller}\ \emph {et~al.}(2017)\citenamefont {Koller},
  \citenamefont {Grotti}, \citenamefont {Vogt}, \citenamefont {Al-Masoudi},
  \citenamefont {D\"{o}rscher}, \citenamefont {H\"{a}fner}, \citenamefont
  {Sterr},\ and\ \citenamefont {Lisdat}}]{Koller2017}%
  \BibitemOpen
  \bibfield  {author} {\bibinfo {author} {\bibfnamefont {S.~B.}\ \bibnamefont
  {Koller}}, \bibinfo {author} {\bibfnamefont {J.}~\bibnamefont {Grotti}},
  \bibinfo {author} {\bibfnamefont {St.}\ \bibnamefont {Vogt}}, \bibinfo
  {author} {\bibfnamefont {A.}~\bibnamefont {Al-Masoudi}}, \bibinfo {author}
  {\bibfnamefont {S.}~\bibnamefont {D\"{o}rscher}}, \bibinfo {author}
  {\bibfnamefont {S.}~\bibnamefont {H\"{a}fner}}, \bibinfo {author}
  {\bibfnamefont {U.}~\bibnamefont {Sterr}}, \ and\ \bibinfo {author}
  {\bibfnamefont {Ch.}\ \bibnamefont {Lisdat}},\ }\bibfield  {title} {\enquote
  {\bibinfo {title} {Transportable optical lattice clock with $7\times10^{-17}$
  uncertainty},}\ }\href {\doibase 10.1103/PhysRevLett.118.073601} {\bibfield
  {journal} {\bibinfo  {journal} {Phys. Rev. Lett.}\ }\textbf {\bibinfo
  {volume} {118}},\ \bibinfo {pages} {073601} (\bibinfo {year}
  {2017})}\BibitemShut {NoStop}%
\bibitem [{\citenamefont {Brewer}\ \emph {et~al.}(2019)\citenamefont {Brewer},
  \citenamefont {Chen}, \citenamefont {Hankin}, \citenamefont {Clements},
  \citenamefont {Chou}, \citenamefont {Wineland}, \citenamefont {Hume},\ and\
  \citenamefont {Leibrandt}}]{Brewer2019}%
  \BibitemOpen
  \bibfield  {author} {\bibinfo {author} {\bibfnamefont {S.~M.}\ \bibnamefont
  {Brewer}}, \bibinfo {author} {\bibfnamefont {J.~S.}\ \bibnamefont {Chen}},
  \bibinfo {author} {\bibfnamefont {A.~M.}\ \bibnamefont {Hankin}}, \bibinfo
  {author} {\bibfnamefont {E.~R.}\ \bibnamefont {Clements}}, \bibinfo {author}
  {\bibfnamefont {C.~W.}\ \bibnamefont {Chou}}, \bibinfo {author}
  {\bibfnamefont {D.~J.}\ \bibnamefont {Wineland}}, \bibinfo {author}
  {\bibfnamefont {D.~B.}\ \bibnamefont {Hume}}, \ and\ \bibinfo {author}
  {\bibfnamefont {D.~R.}\ \bibnamefont {Leibrandt}},\ }\bibfield  {title}
  {\enquote {\bibinfo {title} {$^{27}\mathrm{Al}^{+}$ quantum-logic clock with
  a systematic uncertainty below $10^{-18}$},}\ }\href {\doibase
  10.1103/PhysRevLett.123.033201} {\bibfield  {journal} {\bibinfo  {journal}
  {Phys. Rev. Lett.}\ }\textbf {\bibinfo {volume} {123}},\ \bibinfo {pages}
  {033201} (\bibinfo {year} {2019})}\BibitemShut {NoStop}%
\bibitem [{\citenamefont {Cirac}\ and\ \citenamefont
  {Zoller}(1995)}]{Cirac1995}%
  \BibitemOpen
  \bibfield  {author} {\bibinfo {author} {\bibfnamefont {J.~I.}\ \bibnamefont
  {Cirac}}\ and\ \bibinfo {author} {\bibfnamefont {P.}~\bibnamefont {Zoller}},\
  }\bibfield  {title} {\enquote {\bibinfo {title} {Quantum computations with
  cold trapped ions},}\ }\href {\doibase 10.1103/PhysRevLett.74.4091}
  {\bibfield  {journal} {\bibinfo  {journal} {Phys. Rev. Lett.}\ }\textbf
  {\bibinfo {volume} {74}},\ \bibinfo {pages} {4091--4094} (\bibinfo {year}
  {1995})}\BibitemShut {NoStop}%
\bibitem [{\citenamefont {Abramovici}\ \emph {et~al.}(1992)\citenamefont
  {Abramovici}, \citenamefont {Althouse}, \citenamefont {Drever}, \citenamefont
  {G\"{u}rsel}, \citenamefont {Kawamura}, \citenamefont {Raab}, \citenamefont
  {Shoemaker}, \citenamefont {Sievers}, \citenamefont {Spero}, \citenamefont
  {Thorne}, \citenamefont {Vogt}, \citenamefont {Weiss}, \citenamefont
  {Whitcomb},\ and\ \citenamefont {Zucker}}]{Abramovici1992}%
  \BibitemOpen
  \bibfield  {author} {\bibinfo {author} {\bibfnamefont {A.}~\bibnamefont
  {Abramovici}}, \bibinfo {author} {\bibfnamefont {W.~E.}\ \bibnamefont
  {Althouse}}, \bibinfo {author} {\bibfnamefont {R.~W.~P.}\ \bibnamefont
  {Drever}}, \bibinfo {author} {\bibfnamefont {Y.}~\bibnamefont {G\"{u}rsel}},
  \bibinfo {author} {\bibfnamefont {S.}~\bibnamefont {Kawamura}}, \bibinfo
  {author} {\bibfnamefont {F.~J.}\ \bibnamefont {Raab}}, \bibinfo {author}
  {\bibfnamefont {D.}~\bibnamefont {Shoemaker}}, \bibinfo {author}
  {\bibfnamefont {L.}~\bibnamefont {Sievers}}, \bibinfo {author} {\bibfnamefont
  {R.~E.}\ \bibnamefont {Spero}}, \bibinfo {author} {\bibfnamefont {K.~S.}\
  \bibnamefont {Thorne}}, \bibinfo {author} {\bibfnamefont {R.~E.}\
  \bibnamefont {Vogt}}, \bibinfo {author} {\bibfnamefont {R.}~\bibnamefont
  {Weiss}}, \bibinfo {author} {\bibfnamefont {S.~E.}\ \bibnamefont {Whitcomb}},
  \ and\ \bibinfo {author} {\bibfnamefont {M.~E.}\ \bibnamefont {Zucker}},\
  }\bibfield  {title} {\enquote {\bibinfo {title} {{LIGO}: The laser
  interferometer gravitational-wave observatory},}\ }\href {\doibase
  10.1126/science.256.5055.325} {\bibfield  {journal} {\bibinfo  {journal}
  {Science}\ }\textbf {\bibinfo {volume} {256}},\ \bibinfo {pages} {325--333}
  (\bibinfo {year} {1992})}\BibitemShut {NoStop}%
\bibitem [{\citenamefont {Abbott}\ \emph {et~al.}(2016)\citenamefont {Abbott},
  \citenamefont {Abbott}, \citenamefont {Abbott}, \citenamefont {Abernathy},
  \citenamefont {Acernese}, \citenamefont {Ackley}, \citenamefont {Adams},
  \citenamefont {Adams}, \citenamefont {Addesso}, \citenamefont {Adhikari},
  \citenamefont {Adya}, \citenamefont {Affeldt}, \citenamefont {Agathos},
  \citenamefont {Agatsuma}, \citenamefont {Aggarwal} \emph
  {et~al.}}]{Abbott2016}%
  \BibitemOpen
  \bibfield  {author} {\bibinfo {author} {\bibfnamefont {B.~P.}\ \bibnamefont
  {Abbott}}, \bibinfo {author} {\bibfnamefont {R.}~\bibnamefont {Abbott}},
  \bibinfo {author} {\bibfnamefont {T.~D.}\ \bibnamefont {Abbott}}, \bibinfo
  {author} {\bibfnamefont {M.~R.}\ \bibnamefont {Abernathy}}, \bibinfo {author}
  {\bibfnamefont {F.}~\bibnamefont {Acernese}}, \bibinfo {author}
  {\bibfnamefont {K.}~\bibnamefont {Ackley}}, \bibinfo {author} {\bibfnamefont
  {C.}~\bibnamefont {Adams}}, \bibinfo {author} {\bibfnamefont
  {T.}~\bibnamefont {Adams}}, \bibinfo {author} {\bibfnamefont
  {P.}~\bibnamefont {Addesso}}, \bibinfo {author} {\bibfnamefont {R.~X.}\
  \bibnamefont {Adhikari}}, \bibinfo {author} {\bibfnamefont {V.~B.}\
  \bibnamefont {Adya}}, \bibinfo {author} {\bibfnamefont {C.}~\bibnamefont
  {Affeldt}}, \bibinfo {author} {\bibfnamefont {M.}~\bibnamefont {Agathos}},
  \bibinfo {author} {\bibfnamefont {K.}~\bibnamefont {Agatsuma}}, \bibinfo
  {author} {\bibfnamefont {N.}~\bibnamefont {Aggarwal}},  \emph {et~al.}
  (\bibinfo {collaboration} {{LIGO} Scientific Collaboration and Virgo
  Collaboration}),\ }\bibfield  {title} {\enquote {\bibinfo {title}
  {Observation of gravitational waves from a binary black hole merger},}\
  }\href {\doibase 10.1103/PhysRevLett.116.061102} {\bibfield  {journal}
  {\bibinfo  {journal} {Phys. Rev. Lett.}\ }\textbf {\bibinfo {volume} {116}},\
  \bibinfo {pages} {061102} (\bibinfo {year} {2016})}\BibitemShut {NoStop}%
\bibitem [{\citenamefont {Rafac}\ \emph {et~al.}(2000)\citenamefont {Rafac},
  \citenamefont {Young}, \citenamefont {Beall}, \citenamefont {Itano},
  \citenamefont {Wineland},\ and\ \citenamefont {Bergquist}}]{Rafac2000}%
  \BibitemOpen
  \bibfield  {author} {\bibinfo {author} {\bibfnamefont {R.~J.}\ \bibnamefont
  {Rafac}}, \bibinfo {author} {\bibfnamefont {B.~C.}\ \bibnamefont {Young}},
  \bibinfo {author} {\bibfnamefont {J.~A.}\ \bibnamefont {Beall}}, \bibinfo
  {author} {\bibfnamefont {W.~M.}\ \bibnamefont {Itano}}, \bibinfo {author}
  {\bibfnamefont {D.~J.}\ \bibnamefont {Wineland}}, \ and\ \bibinfo {author}
  {\bibfnamefont {J.~C.}\ \bibnamefont {Bergquist}},\ }\bibfield  {title}
  {\enquote {\bibinfo {title} {Sub-dekahertz ultraviolet spectroscopy of
  ${}^{199}{\mathrm{hg}}^{+}$},}\ }\href {\doibase 10.1103/PhysRevLett.85.2462}
  {\bibfield  {journal} {\bibinfo  {journal} {Phys. Rev. Lett.}\ }\textbf
  {\bibinfo {volume} {85}},\ \bibinfo {pages} {2462--2465} (\bibinfo {year}
  {2000})}\BibitemShut {NoStop}%
\bibitem [{\citenamefont {Lee}\ \emph {et~al.}(2012)\citenamefont {Lee},
  \citenamefont {Chen}, \citenamefont {Li}, \citenamefont {Yang}, \citenamefont
  {Jeon}, \citenamefont {Painter},\ and\ \citenamefont {Vahala}}]{Lee2012}%
  \BibitemOpen
  \bibfield  {author} {\bibinfo {author} {\bibfnamefont {H.}~\bibnamefont
  {Lee}}, \bibinfo {author} {\bibfnamefont {T.}~\bibnamefont {Chen}}, \bibinfo
  {author} {\bibfnamefont {J.}~\bibnamefont {Li}}, \bibinfo {author}
  {\bibfnamefont {K.~Y.}\ \bibnamefont {Yang}}, \bibinfo {author}
  {\bibfnamefont {S.}~\bibnamefont {Jeon}}, \bibinfo {author} {\bibfnamefont
  {O.}~\bibnamefont {Painter}}, \ and\ \bibinfo {author} {\bibfnamefont
  {K.~J.}\ \bibnamefont {Vahala}},\ }\bibfield  {title} {\enquote {\bibinfo
  {title} {Chemically etched ultrahigh-{Q} wedge-resonator on a silicon
  chip},}\ }\href {\doibase 10.1038/nphoton.2012.109} {\bibfield  {journal}
  {\bibinfo  {journal} {Nat. Photon.}\ }\textbf {\bibinfo {volume} {6}},\
  \bibinfo {pages} {369--373} (\bibinfo {year} {2012})}\BibitemShut {NoStop}%
\bibitem [{\citenamefont {Santis}\ \emph {et~al.}(2014)\citenamefont {Santis},
  \citenamefont {Steger}, \citenamefont {Vilenchik}, \citenamefont {Vasilyev},\
  and\ \citenamefont {Yariv}}]{Santis2014}%
  \BibitemOpen
  \bibfield  {author} {\bibinfo {author} {\bibfnamefont {Christos~Theodoros}\
  \bibnamefont {Santis}}, \bibinfo {author} {\bibfnamefont {Scott~T.}\
  \bibnamefont {Steger}}, \bibinfo {author} {\bibfnamefont {Yaakov}\
  \bibnamefont {Vilenchik}}, \bibinfo {author} {\bibfnamefont {Arseny}\
  \bibnamefont {Vasilyev}}, \ and\ \bibinfo {author} {\bibfnamefont {Amnon}\
  \bibnamefont {Yariv}},\ }\bibfield  {title} {\enquote {\bibinfo {title}
  {High-coherence semiconductor lasers based on integral high-q resonators in
  hybrid si/iii-v platforms},}\ }\href {\doibase 10.1073/pnas.1400184111}
  {\bibfield  {journal} {\bibinfo  {journal} {Proceedings of the National
  Academy of Sciences}\ }\textbf {\bibinfo {volume} {111}},\ \bibinfo {pages}
  {2879--2884} (\bibinfo {year} {2014})},\ \Eprint
  {http://arxiv.org/abs/https://www.pnas.org/content/111/8/2879.full.pdf}
  {https://www.pnas.org/content/111/8/2879.full.pdf} \BibitemShut {NoStop}%
\bibitem [{\citenamefont {Morton}\ and\ \citenamefont
  {Morton}(2018)}]{Morton2018}%
  \BibitemOpen
  \bibfield  {author} {\bibinfo {author} {\bibfnamefont {Paul~A.}\ \bibnamefont
  {Morton}}\ and\ \bibinfo {author} {\bibfnamefont {Michael~J.}\ \bibnamefont
  {Morton}},\ }\bibfield  {title} {\enquote {\bibinfo {title} {High-power,
  ultra-low noise hybrid lasers for microwave photonics and optical sensing},}\
  }\href {http://jlt.osa.org/abstract.cfm?URI=jlt-36-21-5048} {\bibfield
  {journal} {\bibinfo  {journal} {J. Lightwave Technol.}\ }\textbf {\bibinfo
  {volume} {36}},\ \bibinfo {pages} {5048--5057} (\bibinfo {year}
  {2018})}\BibitemShut {NoStop}%
\bibitem [{\citenamefont {Gundavarapu}\ \emph {et~al.}(2019)\citenamefont
  {Gundavarapu}, \citenamefont {Brodnik}, \citenamefont {Puckett},
  \citenamefont {Huffman}, \citenamefont {Bose}, \citenamefont {Behunin},
  \citenamefont {Wu}, \citenamefont {Qiu}, \citenamefont {Pinho}, \citenamefont
  {Chauhan}, \citenamefont {Nohava}, \citenamefont {Rakich}, \citenamefont
  {Nelson}, \citenamefont {Salit},\ and\ \citenamefont
  {Blumenthal}}]{Gundavarapu2019}%
  \BibitemOpen
  \bibfield  {author} {\bibinfo {author} {\bibfnamefont {S.}~\bibnamefont
  {Gundavarapu}}, \bibinfo {author} {\bibfnamefont {G.~M.}\ \bibnamefont
  {Brodnik}}, \bibinfo {author} {\bibfnamefont {M.}~\bibnamefont {Puckett}},
  \bibinfo {author} {\bibfnamefont {T.}~\bibnamefont {Huffman}}, \bibinfo
  {author} {\bibfnamefont {D.}~\bibnamefont {Bose}}, \bibinfo {author}
  {\bibfnamefont {R.}~\bibnamefont {Behunin}}, \bibinfo {author} {\bibfnamefont
  {J.}~\bibnamefont {Wu}}, \bibinfo {author} {\bibfnamefont {T.}~\bibnamefont
  {Qiu}}, \bibinfo {author} {\bibfnamefont {C.}~\bibnamefont {Pinho}}, \bibinfo
  {author} {\bibfnamefont {N.}~\bibnamefont {Chauhan}}, \bibinfo {author}
  {\bibfnamefont {J.}~\bibnamefont {Nohava}}, \bibinfo {author} {\bibfnamefont
  {P.~T.}\ \bibnamefont {Rakich}}, \bibinfo {author} {\bibfnamefont {K.~D.}\
  \bibnamefont {Nelson}}, \bibinfo {author} {\bibfnamefont {M.}~\bibnamefont
  {Salit}}, \ and\ \bibinfo {author} {\bibfnamefont {D.~J.}\ \bibnamefont
  {Blumenthal}},\ }\bibfield  {title} {\enquote {\bibinfo {title} {Sub-hertz
  fundamental linewidth photonic integrated brillouin laser},}\ }\href
  {\doibase 10.1038/s41566-018-0313-2} {\bibfield  {journal} {\bibinfo
  {journal} {Nat. Photon.}\ }\textbf {\bibinfo {volume} {13}},\ \bibinfo
  {pages} {60--67} (\bibinfo {year} {2019})}\BibitemShut {NoStop}%
\bibitem [{\citenamefont {Jin}\ \emph {et~al.}(2021)\citenamefont {Jin},
  \citenamefont {Yang}, \citenamefont {Chang}, \citenamefont {Shen},
  \citenamefont {Wang}, \citenamefont {Leal}, \citenamefont {Wu}, \citenamefont
  {Gao}, \citenamefont {Feshali}, \citenamefont {Paniccia}, \citenamefont
  {Vahala},\ and\ \citenamefont {Bowers}}]{Jin2021}%
  \BibitemOpen
  \bibfield  {author} {\bibinfo {author} {\bibfnamefont {W.}~\bibnamefont
  {Jin}}, \bibinfo {author} {\bibfnamefont {Q.}~\bibnamefont {Yang}}, \bibinfo
  {author} {\bibfnamefont {L.}~\bibnamefont {Chang}}, \bibinfo {author}
  {\bibfnamefont {B.}~\bibnamefont {Shen}}, \bibinfo {author} {\bibfnamefont
  {H.}~\bibnamefont {Wang}}, \bibinfo {author} {\bibfnamefont {M.~A.}\
  \bibnamefont {Leal}}, \bibinfo {author} {\bibfnamefont {L.}~\bibnamefont
  {Wu}}, \bibinfo {author} {\bibfnamefont {M.}~\bibnamefont {Gao}}, \bibinfo
  {author} {\bibfnamefont {A.}~\bibnamefont {Feshali}}, \bibinfo {author}
  {\bibfnamefont {M.}~\bibnamefont {Paniccia}}, \bibinfo {author}
  {\bibfnamefont {K.~J.}\ \bibnamefont {Vahala}}, \ and\ \bibinfo {author}
  {\bibfnamefont {J.~E.}\ \bibnamefont {Bowers}},\ }\bibfield  {title}
  {\enquote {\bibinfo {title} {Hertz-linewidth semiconductor lasers using
  cmos-ready ultra-high-q microresonators},}\ }\href {\doibase
  https://doi.org/10.1038/s41566-021-00761-7} {\bibfield  {journal} {\bibinfo
  {journal} {Nat. Photonics}\ }\textbf {\bibinfo {volume} {15}},\ \bibinfo
  {pages} {346--353} (\bibinfo {year} {2021})}\BibitemShut {NoStop}%
\bibitem [{\citenamefont {Debut}\ \emph {et~al.}(2000)\citenamefont {Debut},
  \citenamefont {Randoux},\ and\ \citenamefont {Zemmouri}}]{Debut2000}%
  \BibitemOpen
  \bibfield  {author} {\bibinfo {author} {\bibfnamefont {A.}~\bibnamefont
  {Debut}}, \bibinfo {author} {\bibfnamefont {S.}~\bibnamefont {Randoux}}, \
  and\ \bibinfo {author} {\bibfnamefont {J.}~\bibnamefont {Zemmouri}},\
  }\bibfield  {title} {\enquote {\bibinfo {title} {Linewidth narrowing in
  brillouin lasers: Theoretical analysis},}\ }\href {\doibase
  10.1103/PhysRevA.62.023803} {\bibfield  {journal} {\bibinfo  {journal} {Phys.
  Rev. A}\ }\textbf {\bibinfo {volume} {62}},\ \bibinfo {pages} {023803}
  (\bibinfo {year} {2000})}\BibitemShut {NoStop}%
\bibitem [{\citenamefont {Loh}\ \emph {et~al.}(2015)\citenamefont {Loh},
  \citenamefont {Green}, \citenamefont {Baynes}, \citenamefont {Cole},
  \citenamefont {Quinlan}, \citenamefont {Lee}, \citenamefont {Vahala},
  \citenamefont {Papp},\ and\ \citenamefont {Diddams}}]{Loh2015}%
  \BibitemOpen
  \bibfield  {author} {\bibinfo {author} {\bibfnamefont {W.}~\bibnamefont
  {Loh}}, \bibinfo {author} {\bibfnamefont {A.~A.~S.}\ \bibnamefont {Green}},
  \bibinfo {author} {\bibfnamefont {F.~N.}\ \bibnamefont {Baynes}}, \bibinfo
  {author} {\bibfnamefont {D.~C.}\ \bibnamefont {Cole}}, \bibinfo {author}
  {\bibfnamefont {F.~J.}\ \bibnamefont {Quinlan}}, \bibinfo {author}
  {\bibfnamefont {H.}~\bibnamefont {Lee}}, \bibinfo {author} {\bibfnamefont
  {K.~J.}\ \bibnamefont {Vahala}}, \bibinfo {author} {\bibfnamefont {S.~B.}\
  \bibnamefont {Papp}}, \ and\ \bibinfo {author} {\bibfnamefont {S.~A.}\
  \bibnamefont {Diddams}},\ }\bibfield  {title} {\enquote {\bibinfo {title}
  {Dual-microcavity narrow-linewidth brillouin laser},}\ }\href {\doibase
  10.1364/OPTICA.2.000225} {\bibfield  {journal} {\bibinfo  {journal} {Optica}\
  }\textbf {\bibinfo {volume} {2}},\ \bibinfo {pages} {225--232} (\bibinfo
  {year} {2015})}\BibitemShut {NoStop}%
\bibitem [{\citenamefont {Liu}\ \emph {et~al.}(2021)\citenamefont {Liu},
  \citenamefont {Dallyn}, \citenamefont {Brodnik}, \citenamefont {Harrington},
  \citenamefont {Chauhan}, \citenamefont {Bose}, \citenamefont {Morton},
  \citenamefont {Papp}, \citenamefont {Behunin},\ and\ \citenamefont
  {Blumenthal}}]{Liu2021}%
  \BibitemOpen
  \bibfield  {author} {\bibinfo {author} {\bibfnamefont {K.}~\bibnamefont
  {Liu}}, \bibinfo {author} {\bibfnamefont {J.~H.}\ \bibnamefont {Dallyn}},
  \bibinfo {author} {\bibfnamefont {Isichenko~A.}\ \bibnamefont {Brodnik},
  \bibfnamefont {G.~M.}}, \bibinfo {author} {\bibfnamefont {M.~W.}\
  \bibnamefont {Harrington}}, \bibinfo {author} {\bibfnamefont
  {N.}~\bibnamefont {Chauhan}}, \bibinfo {author} {\bibfnamefont
  {D.}~\bibnamefont {Bose}}, \bibinfo {author} {\bibfnamefont {P.~A.}\
  \bibnamefont {Morton}}, \bibinfo {author} {\bibfnamefont {S.~B.}\
  \bibnamefont {Papp}}, \bibinfo {author} {\bibfnamefont {R.~O.}\ \bibnamefont
  {Behunin}}, \ and\ \bibinfo {author} {\bibfnamefont {D.~J.}\ \bibnamefont
  {Blumenthal}},\ }\bibfield  {title} {\enquote {\bibinfo {title} {Photonic
  circuits for laser stabilization with ultra-low-loss and nonlinear
  resonators},}\ }\href@noop {} {\bibfield  {journal} {\bibinfo  {journal}
  {arXiv:2107.03595}\ } (\bibinfo {year} {2021})}\BibitemShut {NoStop}%
\bibitem [{\citenamefont {Drever}\ \emph {et~al.}(1983)\citenamefont {Drever},
  \citenamefont {Hall}, \citenamefont {Kowalski}, \citenamefont {Hough},
  \citenamefont {Ford}, \citenamefont {Munley},\ and\ \citenamefont
  {Ward}}]{Drever1983}%
  \BibitemOpen
  \bibfield  {author} {\bibinfo {author} {\bibfnamefont {R.~W.~P.}\
  \bibnamefont {Drever}}, \bibinfo {author} {\bibfnamefont {J.~L.}\
  \bibnamefont {Hall}}, \bibinfo {author} {\bibfnamefont {F.~V.}\ \bibnamefont
  {Kowalski}}, \bibinfo {author} {\bibfnamefont {J.}~\bibnamefont {Hough}},
  \bibinfo {author} {\bibfnamefont {G.~M.}\ \bibnamefont {Ford}}, \bibinfo
  {author} {\bibfnamefont {A.~J.}\ \bibnamefont {Munley}}, \ and\ \bibinfo
  {author} {\bibfnamefont {H.}~\bibnamefont {Ward}},\ }\bibfield  {title}
  {\enquote {\bibinfo {title} {Laser phase and frequency stabilization using an
  optical resonator},}\ }\href {\doibase 10.1007/BF00702605} {\bibfield
  {journal} {\bibinfo  {journal} {Appl. Phys. B}\ }\textbf {\bibinfo {volume}
  {31}},\ \bibinfo {pages} {97--105} (\bibinfo {year} {1983})}\BibitemShut
  {NoStop}%
\bibitem [{\citenamefont {Gorodetsky}\ and\ \citenamefont
  {Grudinin}(2004)}]{Gorodetsky2004}%
  \BibitemOpen
  \bibfield  {author} {\bibinfo {author} {\bibfnamefont {Michael~L.}\
  \bibnamefont {Gorodetsky}}\ and\ \bibinfo {author} {\bibfnamefont {Ivan~S.}\
  \bibnamefont {Grudinin}},\ }\bibfield  {title} {\enquote {\bibinfo {title}
  {Fundamental thermal fluctuations in microspheres},}\ }\href {\doibase
  10.1364/JOSAB.21.000697} {\bibfield  {journal} {\bibinfo  {journal} {J. Opt.
  Soc. Am. B}\ }\textbf {\bibinfo {volume} {21}},\ \bibinfo {pages} {697--705}
  (\bibinfo {year} {2004})}\BibitemShut {NoStop}%
\bibitem [{\citenamefont {Notcutt}\ \emph {et~al.}(2006)\citenamefont
  {Notcutt}, \citenamefont {Ma}, \citenamefont {Ludlow}, \citenamefont
  {Foreman}, \citenamefont {Ye},\ and\ \citenamefont {Hall}}]{Notcutt2006}%
  \BibitemOpen
  \bibfield  {author} {\bibinfo {author} {\bibfnamefont {Mark}\ \bibnamefont
  {Notcutt}}, \bibinfo {author} {\bibfnamefont {Long-Sheng}\ \bibnamefont
  {Ma}}, \bibinfo {author} {\bibfnamefont {Andrew~D.}\ \bibnamefont {Ludlow}},
  \bibinfo {author} {\bibfnamefont {Seth~M.}\ \bibnamefont {Foreman}}, \bibinfo
  {author} {\bibfnamefont {Jun}\ \bibnamefont {Ye}}, \ and\ \bibinfo {author}
  {\bibfnamefont {John~L.}\ \bibnamefont {Hall}},\ }\bibfield  {title}
  {\enquote {\bibinfo {title} {Contribution of thermal noise to frequency
  stability of rigid optical cavity via hertz-linewidth lasers},}\ }\href
  {\doibase 10.1103/PhysRevA.73.031804} {\bibfield  {journal} {\bibinfo
  {journal} {Phys. Rev. A}\ }\textbf {\bibinfo {volume} {73}},\ \bibinfo
  {pages} {031804} (\bibinfo {year} {2006})}\BibitemShut {NoStop}%
\bibitem [{\citenamefont {Matsko}\ \emph {et~al.}(2007)\citenamefont {Matsko},
  \citenamefont {Savchenkov}, \citenamefont {Yu},\ and\ \citenamefont
  {Maleki}}]{Matsko2007}%
  \BibitemOpen
  \bibfield  {author} {\bibinfo {author} {\bibfnamefont {Andrey~B.}\
  \bibnamefont {Matsko}}, \bibinfo {author} {\bibfnamefont {Anatoliy~A.}\
  \bibnamefont {Savchenkov}}, \bibinfo {author} {\bibfnamefont {Nan}\
  \bibnamefont {Yu}}, \ and\ \bibinfo {author} {\bibfnamefont {Lute}\
  \bibnamefont {Maleki}},\ }\bibfield  {title} {\enquote {\bibinfo {title}
  {Whispering-gallery-mode resonators as frequency references. i. fundamental
  limitations},}\ }\href {\doibase 10.1364/JOSAB.24.001324} {\bibfield
  {journal} {\bibinfo  {journal} {J. Opt. Soc. Am. B}\ }\textbf {\bibinfo
  {volume} {24}},\ \bibinfo {pages} {1324--1335} (\bibinfo {year}
  {2007})}\BibitemShut {NoStop}%
\bibitem [{\citenamefont {Webster}\ \emph {et~al.}(2008)\citenamefont
  {Webster}, \citenamefont {Oxborrow}, \citenamefont {Pugla}, \citenamefont
  {Millo},\ and\ \citenamefont {Gill}}]{Webster2008}%
  \BibitemOpen
  \bibfield  {author} {\bibinfo {author} {\bibfnamefont {S.~A.}\ \bibnamefont
  {Webster}}, \bibinfo {author} {\bibfnamefont {M.}~\bibnamefont {Oxborrow}},
  \bibinfo {author} {\bibfnamefont {S.}~\bibnamefont {Pugla}}, \bibinfo
  {author} {\bibfnamefont {J.}~\bibnamefont {Millo}}, \ and\ \bibinfo {author}
  {\bibfnamefont {P.}~\bibnamefont {Gill}},\ }\bibfield  {title} {\enquote
  {\bibinfo {title} {Thermal-noise-limited optical cavity},}\ }\href {\doibase
  10.1103/PhysRevA.77.033847} {\bibfield  {journal} {\bibinfo  {journal} {Phys.
  Rev. A}\ }\textbf {\bibinfo {volume} {77}},\ \bibinfo {pages} {033847}
  (\bibinfo {year} {2008})}\BibitemShut {NoStop}%
\bibitem [{\citenamefont {Sun}\ \emph {et~al.}(2017)\citenamefont {Sun},
  \citenamefont {Luo}, \citenamefont {Zhang},\ and\ \citenamefont
  {Lin}}]{Sun2017}%
  \BibitemOpen
  \bibfield  {author} {\bibinfo {author} {\bibfnamefont {Xuan}\ \bibnamefont
  {Sun}}, \bibinfo {author} {\bibfnamefont {Rui}\ \bibnamefont {Luo}}, \bibinfo
  {author} {\bibfnamefont {Xi-Cheng}\ \bibnamefont {Zhang}}, \ and\ \bibinfo
  {author} {\bibfnamefont {Qiang}\ \bibnamefont {Lin}},\ }\bibfield  {title}
  {\enquote {\bibinfo {title} {Squeezing the fundamental temperature
  fluctuations of a high-$q$ microresonator},}\ }\href {\doibase
  10.1103/PhysRevA.95.023822} {\bibfield  {journal} {\bibinfo  {journal} {Phys.
  Rev. A}\ }\textbf {\bibinfo {volume} {95}},\ \bibinfo {pages} {023822}
  (\bibinfo {year} {2017})}\BibitemShut {NoStop}%
\bibitem [{\citenamefont {Lim}\ \emph {et~al.}(2019)\citenamefont {Lim},
  \citenamefont {Liang}, \citenamefont {Savchenkov}, \citenamefont {Matsko},
  \citenamefont {Maleki},\ and\ \citenamefont {Wong}}]{Lim2019}%
  \BibitemOpen
  \bibfield  {author} {\bibinfo {author} {\bibfnamefont {J.}~\bibnamefont
  {Lim}}, \bibinfo {author} {\bibfnamefont {W.}~\bibnamefont {Liang}}, \bibinfo
  {author} {\bibfnamefont {A.~A.}\ \bibnamefont {Savchenkov}}, \bibinfo
  {author} {\bibfnamefont {A.~B.}\ \bibnamefont {Matsko}}, \bibinfo {author}
  {\bibfnamefont {L.}~\bibnamefont {Maleki}}, \ and\ \bibinfo {author}
  {\bibfnamefont {C.~W.}\ \bibnamefont {Wong}},\ }\bibfield  {title} {\enquote
  {\bibinfo {title} {Probing $10$ $\mu\mathrm{K}$ stability and residual drifts
  in the cross-polarization dual-mode stabilization of single-crystal
  ultrahigh-{Q} optical resonators},}\ }\href {\doibase
  10.1038/s41377-018-0109-7} {\bibfield  {journal} {\bibinfo  {journal} {Light
  Sci. Appl.}\ }\textbf {\bibinfo {volume} {8}},\ \bibinfo {pages} {1}
  (\bibinfo {year} {2019})}\BibitemShut {NoStop}%
\bibitem [{\citenamefont {Huang}\ \emph {et~al.}(2019)\citenamefont {Huang},
  \citenamefont {Lucas}, \citenamefont {Liu}, \citenamefont {Raja},
  \citenamefont {Lihachev}, \citenamefont {Gorodetsky}, \citenamefont
  {Engelsen},\ and\ \citenamefont {Kippenberg}}]{Huang2019}%
  \BibitemOpen
  \bibfield  {author} {\bibinfo {author} {\bibfnamefont {Guanhao}\ \bibnamefont
  {Huang}}, \bibinfo {author} {\bibfnamefont {Erwan}\ \bibnamefont {Lucas}},
  \bibinfo {author} {\bibfnamefont {Junqiu}\ \bibnamefont {Liu}}, \bibinfo
  {author} {\bibfnamefont {Arslan~S.}\ \bibnamefont {Raja}}, \bibinfo {author}
  {\bibfnamefont {Grigory}\ \bibnamefont {Lihachev}}, \bibinfo {author}
  {\bibfnamefont {Michael~L.}\ \bibnamefont {Gorodetsky}}, \bibinfo {author}
  {\bibfnamefont {Nils~J.}\ \bibnamefont {Engelsen}}, \ and\ \bibinfo {author}
  {\bibfnamefont {Tobias~J.}\ \bibnamefont {Kippenberg}},\ }\bibfield  {title}
  {\enquote {\bibinfo {title} {Thermorefractive noise in silicon-nitride
  microresonators},}\ }\href {\doibase 10.1103/PhysRevA.99.061801} {\bibfield
  {journal} {\bibinfo  {journal} {Phys. Rev. A}\ }\textbf {\bibinfo {volume}
  {99}},\ \bibinfo {pages} {061801} (\bibinfo {year} {2019})}\BibitemShut
  {NoStop}%
\bibitem [{\citenamefont {Panuski}\ \emph {et~al.}(2020)\citenamefont
  {Panuski}, \citenamefont {Englund},\ and\ \citenamefont
  {Hamerly}}]{Panuski2020}%
  \BibitemOpen
  \bibfield  {author} {\bibinfo {author} {\bibfnamefont {Christopher}\
  \bibnamefont {Panuski}}, \bibinfo {author} {\bibfnamefont {Dirk}\
  \bibnamefont {Englund}}, \ and\ \bibinfo {author} {\bibfnamefont {Ryan}\
  \bibnamefont {Hamerly}},\ }\bibfield  {title} {\enquote {\bibinfo {title}
  {Fundamental thermal noise limits for optical microcavities},}\ }\href
  {\doibase 10.1103/PhysRevX.10.041046} {\bibfield  {journal} {\bibinfo
  {journal} {Phys. Rev. X}\ }\textbf {\bibinfo {volume} {10}},\ \bibinfo
  {pages} {041046} (\bibinfo {year} {2020})}\BibitemShut {NoStop}%
\bibitem [{\citenamefont {Li}\ \emph {et~al.}(2021)\citenamefont {Li},
  \citenamefont {Jin}, \citenamefont {Wu}, \citenamefont {Chang}, \citenamefont
  {Wang}, \citenamefont {Shen}, \citenamefont {Yuan}, \citenamefont {Feshali},
  \citenamefont {Paniccia}, \citenamefont {Vahala},\ and\ \citenamefont
  {Bowers}}]{Li2021}%
  \BibitemOpen
  \bibfield  {author} {\bibinfo {author} {\bibfnamefont {B.}~\bibnamefont
  {Li}}, \bibinfo {author} {\bibfnamefont {W}~\bibnamefont {Jin}}, \bibinfo
  {author} {\bibfnamefont {L.}~\bibnamefont {Wu}}, \bibinfo {author}
  {\bibfnamefont {L.}~\bibnamefont {Chang}}, \bibinfo {author} {\bibfnamefont
  {H.}~\bibnamefont {Wang}}, \bibinfo {author} {\bibfnamefont {B.}~\bibnamefont
  {Shen}}, \bibinfo {author} {\bibfnamefont {Z.}~\bibnamefont {Yuan}}, \bibinfo
  {author} {\bibfnamefont {A.}~\bibnamefont {Feshali}}, \bibinfo {author}
  {\bibfnamefont {M.}~\bibnamefont {Paniccia}}, \bibinfo {author}
  {\bibfnamefont {K.~J.}\ \bibnamefont {Vahala}}, \ and\ \bibinfo {author}
  {\bibfnamefont {J.~E.}\ \bibnamefont {Bowers}},\ }\bibfield  {title}
  {\enquote {\bibinfo {title} {Reaching fiber-laser coherence in integrated
  photonics},}\ }\href {\doibase 10.1364/OL.439720} {\bibfield  {journal}
  {\bibinfo  {journal} {Opt. Lett.}\ }\textbf {\bibinfo {volume} {46}},\
  \bibinfo {pages} {5201--5204} (\bibinfo {year} {2021})}\BibitemShut {NoStop}%
\bibitem [{\citenamefont {Grudinin}\ \emph {et~al.}(2009)\citenamefont
  {Grudinin}, \citenamefont {Matsko},\ and\ \citenamefont
  {Maleki}}]{Grudinin2009}%
  \BibitemOpen
  \bibfield  {author} {\bibinfo {author} {\bibfnamefont {I.~S.}\ \bibnamefont
  {Grudinin}}, \bibinfo {author} {\bibfnamefont {A.~B.}\ \bibnamefont
  {Matsko}}, \ and\ \bibinfo {author} {\bibfnamefont {L.}~\bibnamefont
  {Maleki}},\ }\bibfield  {title} {\enquote {\bibinfo {title} {Brillouin lasing
  with a $\mathrm{CaF}_{2}$ whispering gallery mode resonator},}\ }\href
  {\doibase 10.1103/PhysRevLett.102.043902} {\bibfield  {journal} {\bibinfo
  {journal} {Phys. Rev. Lett.}\ }\textbf {\bibinfo {volume} {102}},\ \bibinfo
  {pages} {043902} (\bibinfo {year} {2009})}\BibitemShut {NoStop}%
\bibitem [{\citenamefont {Kabakova}\ \emph {et~al.}(2013)\citenamefont
  {Kabakova}, \citenamefont {Pant}, \citenamefont {Choi}, \citenamefont
  {Debbarma}, \citenamefont {Luther-Davies}, \citenamefont {Madden},\ and\
  \citenamefont {Eggleton}}]{Kabakova2013}%
  \BibitemOpen
  \bibfield  {author} {\bibinfo {author} {\bibfnamefont {I.~V.}\ \bibnamefont
  {Kabakova}}, \bibinfo {author} {\bibfnamefont {R.}~\bibnamefont {Pant}},
  \bibinfo {author} {\bibfnamefont {D.~Y.}\ \bibnamefont {Choi}}, \bibinfo
  {author} {\bibfnamefont {S.}~\bibnamefont {Debbarma}}, \bibinfo {author}
  {\bibfnamefont {B.}~\bibnamefont {Luther-Davies}}, \bibinfo {author}
  {\bibfnamefont {S.~J.}\ \bibnamefont {Madden}}, \ and\ \bibinfo {author}
  {\bibfnamefont {B.~J.}\ \bibnamefont {Eggleton}},\ }\bibfield  {title}
  {\enquote {\bibinfo {title} {Narrow linewidth brillouin laser based on
  chalcogenide photonic chip},}\ }\href {\doibase 10.1364/OL.38.003208}
  {\bibfield  {journal} {\bibinfo  {journal} {Opt. Lett.}\ }\textbf {\bibinfo
  {volume} {38}},\ \bibinfo {pages} {3208--3211} (\bibinfo {year}
  {2013})}\BibitemShut {NoStop}%
\bibitem [{\citenamefont {Otterstrom}\ \emph {et~al.}(2018)\citenamefont
  {Otterstrom}, \citenamefont {Behunin}, \citenamefont {Kittlaus},
  \citenamefont {Wang},\ and\ \citenamefont {Rakich}}]{Otterstrom2018}%
  \BibitemOpen
  \bibfield  {author} {\bibinfo {author} {\bibfnamefont {N.~T.}\ \bibnamefont
  {Otterstrom}}, \bibinfo {author} {\bibfnamefont {R.~O.}\ \bibnamefont
  {Behunin}}, \bibinfo {author} {\bibfnamefont {E.~A.}\ \bibnamefont
  {Kittlaus}}, \bibinfo {author} {\bibfnamefont {Z.}~\bibnamefont {Wang}}, \
  and\ \bibinfo {author} {\bibfnamefont {P.~T.}\ \bibnamefont {Rakich}},\
  }\bibfield  {title} {\enquote {\bibinfo {title} {A silicon brillouin
  laser},}\ }\href {\doibase 10.1126/science.aar6113} {\bibfield  {journal}
  {\bibinfo  {journal} {Science}\ }\textbf {\bibinfo {volume} {360}},\ \bibinfo
  {pages} {1113--1116} (\bibinfo {year} {2018})}\BibitemShut {NoStop}%
\bibitem [{\citenamefont {Geng}\ \emph {et~al.}(2006)\citenamefont {Geng},
  \citenamefont {Staines}, \citenamefont {Wang}, \citenamefont {Zong},
  \citenamefont {Blake},\ and\ \citenamefont {Jiang}}]{Geng2006}%
  \BibitemOpen
  \bibfield  {author} {\bibinfo {author} {\bibfnamefont {J.}~\bibnamefont
  {Geng}}, \bibinfo {author} {\bibfnamefont {S.}~\bibnamefont {Staines}},
  \bibinfo {author} {\bibfnamefont {Z.}~\bibnamefont {Wang}}, \bibinfo {author}
  {\bibfnamefont {J.}~\bibnamefont {Zong}}, \bibinfo {author} {\bibfnamefont
  {M.}~\bibnamefont {Blake}}, \ and\ \bibinfo {author} {\bibfnamefont
  {S.}~\bibnamefont {Jiang}},\ }\bibfield  {title} {\enquote {\bibinfo {title}
  {Highly stable low-noise \uppercase{B}rillouin fiber laser with ultranarrow
  spectral linewidth},}\ }\href {\doibase 10.1109/LPT.2006.881145} {\bibfield
  {journal} {\bibinfo  {journal} {IEEE Photonics Technology Letters}\ }\textbf
  {\bibinfo {volume} {18}},\ \bibinfo {pages} {1813--1815} (\bibinfo {year}
  {2006})}\BibitemShut {NoStop}%
\bibitem [{\citenamefont {Shee}\ \emph {et~al.}(2011)\citenamefont {Shee},
  \citenamefont {Al-Mansoori}, \citenamefont {Ismail}, \citenamefont {Hitam},\
  and\ \citenamefont {Mahdi}}]{Shee2011}%
  \BibitemOpen
  \bibfield  {author} {\bibinfo {author} {\bibfnamefont {Y.~G.}\ \bibnamefont
  {Shee}}, \bibinfo {author} {\bibfnamefont {M.~H.}\ \bibnamefont
  {Al-Mansoori}}, \bibinfo {author} {\bibfnamefont {A.}~\bibnamefont {Ismail}},
  \bibinfo {author} {\bibfnamefont {S.}~\bibnamefont {Hitam}}, \ and\ \bibinfo
  {author} {\bibfnamefont {M.~A.}\ \bibnamefont {Mahdi}},\ }\bibfield  {title}
  {\enquote {\bibinfo {title} {Multiwavelength \uppercase{B}rillouin-erbium
  fiber laser with double-\uppercase{B}rillouin-frequency spacing},}\ }\href
  {\doibase 10.1364/OE.19.001699} {\bibfield  {journal} {\bibinfo  {journal}
  {Opt. Express}\ }\textbf {\bibinfo {volume} {19}},\ \bibinfo {pages}
  {1699--1706} (\bibinfo {year} {2011})}\BibitemShut {NoStop}%
\bibitem [{\citenamefont {Tow}\ \emph {et~al.}(2012)\citenamefont {Tow},
  \citenamefont {L\'{e}guillon}, \citenamefont {Fresnel}, \citenamefont
  {Besnard}, \citenamefont {Brilland}, \citenamefont {M\'{e}chin},
  \citenamefont {Tr\'{e}goat}, \citenamefont {Troles},\ and\ \citenamefont
  {Toupin}}]{Tow2012}%
  \BibitemOpen
  \bibfield  {author} {\bibinfo {author} {\bibfnamefont {K.~H.}\ \bibnamefont
  {Tow}}, \bibinfo {author} {\bibfnamefont {Y.}~\bibnamefont {L\'{e}guillon}},
  \bibinfo {author} {\bibfnamefont {S.}~\bibnamefont {Fresnel}}, \bibinfo
  {author} {\bibfnamefont {P.}~\bibnamefont {Besnard}}, \bibinfo {author}
  {\bibfnamefont {L.}~\bibnamefont {Brilland}}, \bibinfo {author}
  {\bibfnamefont {D.}~\bibnamefont {M\'{e}chin}}, \bibinfo {author}
  {\bibfnamefont {D.}~\bibnamefont {Tr\'{e}goat}}, \bibinfo {author}
  {\bibfnamefont {J.}~\bibnamefont {Troles}}, \ and\ \bibinfo {author}
  {\bibfnamefont {P.}~\bibnamefont {Toupin}},\ }\bibfield  {title} {\enquote
  {\bibinfo {title} {Linewidth-narrowing and intensity noise reduction of the
  $2^{nd}$ order stokes component of a low threshold \uppercase{B}rillouin
  laser made of \uppercase{G}e$_{10}$\uppercase{A}s$_{22}$\uppercase{S}e$_{68}$
  chalcogenide fiber},}\ }\href {\doibase 10.1364/OE.20.00B104} {\bibfield
  {journal} {\bibinfo  {journal} {Opt. Express}\ }\textbf {\bibinfo {volume}
  {20}},\ \bibinfo {pages} {B104--B109} (\bibinfo {year} {2012})}\BibitemShut
  {NoStop}%
\bibitem [{\citenamefont {Loh}\ \emph {et~al.}(2019)\citenamefont {Loh},
  \citenamefont {Yegnanarayanan}, \citenamefont {O'Donnell},\ and\
  \citenamefont {Juodawlkis}}]{Loh2019}%
  \BibitemOpen
  \bibfield  {author} {\bibinfo {author} {\bibfnamefont {W.}~\bibnamefont
  {Loh}}, \bibinfo {author} {\bibfnamefont {S.}~\bibnamefont {Yegnanarayanan}},
  \bibinfo {author} {\bibfnamefont {F.}~\bibnamefont {O'Donnell}}, \ and\
  \bibinfo {author} {\bibfnamefont {P.~W.}\ \bibnamefont {Juodawlkis}},\
  }\bibfield  {title} {\enquote {\bibinfo {title} {Ultra-narrow linewidth
  brillouin laser with nanokelvin temperature self-referencing},}\ }\href
  {\doibase 10.1364/OPTICA.6.000152} {\bibfield  {journal} {\bibinfo  {journal}
  {Optica}\ }\textbf {\bibinfo {volume} {6}},\ \bibinfo {pages} {152--159}
  (\bibinfo {year} {2019})}\BibitemShut {NoStop}%
\bibitem [{\citenamefont {Loh}\ \emph {et~al.}(2020)\citenamefont {Loh},
  \citenamefont {Stuart}, \citenamefont {Reens}, \citenamefont {Bruzewicz},
  \citenamefont {Braje}, \citenamefont {Chiaverini}, \citenamefont {Sage},\
  and\ \citenamefont {McConnell}}]{Loh2020}%
  \BibitemOpen
  \bibfield  {author} {\bibinfo {author} {\bibfnamefont {W.}~\bibnamefont
  {Loh}}, \bibinfo {author} {\bibfnamefont {J.}~\bibnamefont {Stuart}},
  \bibinfo {author} {\bibfnamefont {D.}~\bibnamefont {Reens}}, \bibinfo
  {author} {\bibfnamefont {C.~D.}\ \bibnamefont {Bruzewicz}}, \bibinfo {author}
  {\bibfnamefont {D.}~\bibnamefont {Braje}}, \bibinfo {author} {\bibfnamefont
  {Juodawlkis P.~W.}\ \bibnamefont {Chiaverini}, \bibfnamefont {J.}}, \bibinfo
  {author} {\bibfnamefont {J.~M.}\ \bibnamefont {Sage}}, \ and\ \bibinfo
  {author} {\bibfnamefont {R.}~\bibnamefont {McConnell}},\ }\bibfield  {title}
  {\enquote {\bibinfo {title} {Operation of an optical atomic clock with a
  brillouin laser subsystem},}\ }\href {\doibase
  https://doi.org/10.1038/s41586-020-2981-6} {\bibfield  {journal} {\bibinfo
  {journal} {Nature}\ }\textbf {\bibinfo {volume} {588}},\ \bibinfo {pages}
  {244--249} (\bibinfo {year} {2020})}\BibitemShut {NoStop}%
\bibitem [{\citenamefont {Bauters}\ \emph {et~al.}(2011)\citenamefont
  {Bauters}, \citenamefont {Heck}, \citenamefont {John}, \citenamefont {Dai},
  \citenamefont {Tien}, \citenamefont {Barton}, \citenamefont {Leinse},
  \citenamefont {Heideman}, \citenamefont {Blumenthal},\ and\ \citenamefont
  {Bowers}}]{Bauters2011}%
  \BibitemOpen
  \bibfield  {author} {\bibinfo {author} {\bibfnamefont {Jared~F.}\
  \bibnamefont {Bauters}}, \bibinfo {author} {\bibfnamefont {Martijn J.~R.}\
  \bibnamefont {Heck}}, \bibinfo {author} {\bibfnamefont {Demis}\ \bibnamefont
  {John}}, \bibinfo {author} {\bibfnamefont {Daoxin}\ \bibnamefont {Dai}},
  \bibinfo {author} {\bibfnamefont {Ming-Chun}\ \bibnamefont {Tien}}, \bibinfo
  {author} {\bibfnamefont {Jonathon~S.}\ \bibnamefont {Barton}}, \bibinfo
  {author} {\bibfnamefont {Arne}\ \bibnamefont {Leinse}}, \bibinfo {author}
  {\bibfnamefont {Ren\'{e}~G.}\ \bibnamefont {Heideman}}, \bibinfo {author}
  {\bibfnamefont {Daniel~J.}\ \bibnamefont {Blumenthal}}, \ and\ \bibinfo
  {author} {\bibfnamefont {John~E.}\ \bibnamefont {Bowers}},\ }\bibfield
  {title} {\enquote {\bibinfo {title} {Ultra-low-loss high-aspect-ratio si3n4
  waveguides},}\ }\href {\doibase 10.1364/OE.19.003163} {\bibfield  {journal}
  {\bibinfo  {journal} {Opt. Express}\ }\textbf {\bibinfo {volume} {19}},\
  \bibinfo {pages} {3163--3174} (\bibinfo {year} {2011})}\BibitemShut {NoStop}%
\bibitem [{\citenamefont {Drake}\ \emph {et~al.}(2020)\citenamefont {Drake},
  \citenamefont {Stone}, \citenamefont {Briles},\ and\ \citenamefont
  {Papp}}]{Drake2020}%
  \BibitemOpen
  \bibfield  {author} {\bibinfo {author} {\bibfnamefont {T.~E.}\ \bibnamefont
  {Drake}}, \bibinfo {author} {\bibfnamefont {J.~R.}\ \bibnamefont {Stone}},
  \bibinfo {author} {\bibfnamefont {T.~C.}\ \bibnamefont {Briles}}, \ and\
  \bibinfo {author} {\bibfnamefont {S.}~\bibnamefont {Papp}},\ }\bibfield
  {title} {\enquote {\bibinfo {title} {Thermal decoherence and laser cooling of
  kerr microresonator solitons},}\ }\href {\doibase
  https://doi.org/10.1038/s41566-020-0651-8} {\bibfield  {journal} {\bibinfo
  {journal} {Nat. Photonics}\ }\textbf {\bibinfo {volume} {14}},\ \bibinfo
  {pages} {480--485} (\bibinfo {year} {2020})}\BibitemShut {NoStop}%
\bibitem [{\citenamefont {Hjelme}\ \emph {et~al.}(1991)\citenamefont {Hjelme},
  \citenamefont {Mickelson},\ and\ \citenamefont {Beausoleil}}]{Hjelme1991}%
  \BibitemOpen
  \bibfield  {author} {\bibinfo {author} {\bibfnamefont {D.~R.}\ \bibnamefont
  {Hjelme}}, \bibinfo {author} {\bibfnamefont {A.~R.}\ \bibnamefont
  {Mickelson}}, \ and\ \bibinfo {author} {\bibfnamefont {R.~G.}\ \bibnamefont
  {Beausoleil}},\ }\bibfield  {title} {\enquote {\bibinfo {title}
  {Semiconductor laser stabilization by external optical feedback},}\ }\href
  {\doibase 10.1109/3.81333} {\bibfield  {journal} {\bibinfo  {journal} {IEEE
  Journal of Quantum Electronics}\ }\textbf {\bibinfo {volume} {27}},\ \bibinfo
  {pages} {352--372} (\bibinfo {year} {1991})}\BibitemShut {NoStop}%
\bibitem [{\citenamefont {Llopis}\ \emph {et~al.}(2011)\citenamefont {Llopis},
  \citenamefont {Merrer}, \citenamefont {Brahimi}, \citenamefont {Saleh},\ and\
  \citenamefont {Lacroix}}]{Llopis2011}%
  \BibitemOpen
  \bibfield  {author} {\bibinfo {author} {\bibfnamefont {O.}~\bibnamefont
  {Llopis}}, \bibinfo {author} {\bibfnamefont {P.~H.}\ \bibnamefont {Merrer}},
  \bibinfo {author} {\bibfnamefont {H.}~\bibnamefont {Brahimi}}, \bibinfo
  {author} {\bibfnamefont {K.}~\bibnamefont {Saleh}}, \ and\ \bibinfo {author}
  {\bibfnamefont {P.}~\bibnamefont {Lacroix}},\ }\bibfield  {title} {\enquote
  {\bibinfo {title} {Phase noise measurement of a narrow linewidth cw laser
  using delay line approaches},}\ }\href {\doibase 10.1364/OL.36.002713}
  {\bibfield  {journal} {\bibinfo  {journal} {Opt. Lett.}\ }\textbf {\bibinfo
  {volume} {36}},\ \bibinfo {pages} {2713--2715} (\bibinfo {year}
  {2011})}\BibitemShut {NoStop}%
\end{thebibliography}%

%%%%%%%%%%%%%%%%%%%%%%%%%%%%%%%%%%%%%%%%%%%%%%%%%%%%%%%%%%%%%%%%

\end{document}